\documentclass[preprint,sort&compress,12pt]{elsarticle}

\usepackage{amssymb}
\usepackage{amsthm}
\usepackage{amsmath}
\usepackage{mathrsfs}
\usepackage{mathtools}
\usepackage{graphicx}
\usepackage{subfig}
\usepackage{floatrow}
\usepackage{diagbox}

\usepackage{caption}

\usepackage{algorithm}
\usepackage{algpseudocode}
\usepackage{array}
\usepackage{multirow}
\usepackage{enumerate}
\usepackage{lineno}
\usepackage{fullpage}
\usepackage{xcolor}
\usepackage[colorinlistoftodos]{todonotes}
\usepackage{bbm}
\usepackage{listings}
\usepackage[colorlinks=true]{hyperref}
\usepackage{url}
\usepackage{siunitx}

\graphicspath{ {./figs/} }
\biboptions{comma,square}

\graphicspath{ {./figs/} }

\journal{Elsevier}

\newcommand{\norm}[1]{\left\lVert#1\right\rVert}
\let\oldequation\equation
\let\oldendequation\endequation

\renewenvironment{equation}
  {\linenomathNonumbers\oldequation}
  {\oldendequation\endlinenomath}
\begin{document}
\begin{frontmatter}

 \title{Surrogate Modeling for Fluid Flows Based on Physics-Constrained Deep Learning Without Simulation Data}



\author[ndAME,ndCICS]{Luning Sun}
\author[ndAME,ndCICS]{Han Gao}
\author[umich]{Shaowu Pan}
\author[ndAME,ndCICS]{Jian-Xun Wang\corref{corxh}}

\address[ndAME]{Department of Aerospace and Mechanical Engineering, University of Notre Dame, Notre Dame, IN}
\address[ndCICS]{Center for Informatics and Computational Science, University of Notre Dame, Notre Dame, IN}
\address[umich]{Department of Aerospace Engineering, University of Michigan, Ann Arbor, MI}
\cortext[corxh]{Corresponding author. Tel: +1 540 315 6512}
\ead{jwang33@nd.edu}

\begin{abstract}
Numerical simulations on fluid dynamics problems primarily rely on spatially or/and temporally discretization of the governing equation using polynomials into a finite-dimensional algebraic system. Due to the multi-scale nature of the physics and sensitivity from meshing a complicated geometry, such process can be computational prohibitive for most real-time applications (e.g., clinical diagnosis and surgery planning) and many-query analyses (e.g., optimization design and uncertainty quantification). Therefore, developing a cost-effective surrogate model is of great practical significance. Deep learning (DL) has shown new promises for surrogate modeling due to its capability of handling strong nonlinearity and high dimensionality. However, the off-the-shelf DL architectures, success of which heavily relies on the large amount of training data and interpolatory nature of the problem, fail to operate when the data becomes sparse. Unfortunately, data is often insufficient in most parametric fluid dynamics problems since each data point in the parameter space requires an expensive numerical simulation based on the first principle, e.g., Naiver--Stokes equations. In this paper, we provide a physics-constrained DL approach for surrogate modeling of fluid flows \emph{without} relying on any simulation data. Specifically, a structured deep neural network (DNN) architecture is devised to enforce the initial and boundary conditions, and the governing partial differential equations (i.e., Navier--Stokes equations) are incorporated into the loss of the DNN to drive the training. Numerical experiments are conducted on a number of internal flows relevant to hemodynamics applications, and the forward propagation of uncertainties in fluid properties and domain geometry is studied as well. The results show excellent agreement on the flow field and forward-propagated uncertainties between the DL surrogate approximations and the first-principle numerical simulations.       
\end{abstract}

\end{frontmatter}

\section{Introduction}
\label{sec:intro}
Complex fluids are ubiquitous in natural and industrial processes, and accurately simulating the fluid flows is indispensable in many disciplines, e.g., aerospace, civil, and biomedical engineering. A fluid system is typically governed by the Navier--Stokes equations, which is a highly nonlinear partial differential equation (PDE) system. Numerical simulation on fluid dynamics problems primarily relies on solving the PDE systems in a discretized form using, e.g., finite difference (FD), finite volume (FV), or finite element (FE) methods, which is known as the computational fluid dynamics (CFD) approach~\cite{anderson1995computational}. However, CFD simulations are often computationally cumbersome, especially for the flows with turbulence and complex geometries. Moreover, mesh generation also usually incurs a huge burden, in particular when moving boundary or large geometric variation is considered. The considerable computational expense greatly limits the use of principled CFD model to \emph{real-time predictions} and \emph{many-query analysis}, which are highly relevant to many scientific problems and real-life applications, e.g., timely clinical diagnosis and surgery planning for cardiovascular diseases, optimization design of aircraft with large parameter variations, and forward/inverse uncertainty quantification (UQ) in high-consequence systems. As an alternative, a cost-effective surrogate model is a computationally feasible way to tackle the aforementioned challenges.

A surrogate model only approximates the input-output relation of a system, which can be evaluated efficiently. Namely, given input parameters, e.g., initial/boundary/operational conditions, the quantities of interest (QoIs), such as velocity, pressure, shear stress, and their integrals can be obtained rapidly without conducting the principled CFD simulations. The existing surrogate modeling approaches can be roughly categorized into two classes: projection-based reduced order models (ROMs) and data-fit models~\cite{benner2015survey}. In projection-based ROMs, a reduced basis is extracted from the simulation data using an unsupervised learning technique, e.g., proper orthogonal decomposition (POD, also known as principal component analysis)~\cite{benner2017model}, and the full-order PDE operator is projected onto the subspace spanned by the reduced basis. As a result, the degrees of freedom of the system can be significantly reduced, and meanwhile, the underlying structure of the full-order model can be retained to a certain extent. Although holding some promises, the current projection-based ROM techniques have had limited impact on complex fluid dynamic problems mainly because of the stability and robustness issues~\cite{lassila2014model,huang2018challenges}. Moreover, projection-based ROMs are highly code-intrusive and their speedup potential is limited when strong nonlinearity exists~\cite{chaturantabut2010nonlinear} though several remedies such as sparse sampling~\cite{peherstorfer2018model,peherstorfer2018stabilizing} exist. Another way to enable rapid simulations is to build a data-fit model, where a response surface of the system is learned from the simulation data in a supervised manner. Namely, a deterministic or probabilistic input-output mapping is constructed using, e.g., polynomial basis functions~\cite{ren2010finite}, radial basis functions~\cite{regis2007stochastic}, Gaussian process (GP)~\cite{kennedy2000predicting,atkinson2019structured}, and stochastic polynomial chaos expansion (PCE)~\cite{xiu2002wiener,najm2009uncertainty,yang2017general}, among others. All these models are built upon the CFD solutions of selected collocation points in parameter space without the need to modify the codes of the CFD solver. Because of the non-intrusive feature and ease of implementation, data-fit surrogates have been used for a wide range of forward and inverse uncertainty quantification (UQ) problems in fluid dynamics~\cite{najm2009uncertainty,le2010spectral,wang2018propagation}. However, traditional data-fit models have a hard time handling the problems with strong nonlinearities and high dimensionality. Deep learning (DL), in particular, the deep neural network (DNN) has become a popular surrogate modeling approach and has shown great potential to deal with high-dimensional nonlinear UQ problems~\cite{zhu2018bayesian,tripathy2018deep,mo2018deep}. It has been shown that DNN as a universal function approximator~\cite{scarselli1998universal} can overcome the curse of dimensionality in certain problems~\cite{hutzenthaler2018overcoming,grohs2018proof,hutzenthaler2019proof}. In broader scientific computing and physical modeling communities, machine learning (ML) has been receiving a lot of attentions~\cite{lee2018basic,carleo2017solving,wang2017physics,ling2016machine,shanahan2018machine,king2018deep,brunton2019data}. However, the tremendous success of DL in the computer science, witnessed in areas of computer vision and image recognition~\cite{lecun2015deep}, can be mainly attributed to the \emph{availability of large-scale labeled data} (i.e., ``big data") and the \emph{interpolatory nature of their problems}. Unfortunately, labeled data for surrogate modeling of fluid systems are often \emph{sparse and could be noisy}, since they are obtained from either principled CFD simulations or experimental observations, both of which are expensive to obtain. Therefore, in such ``small data" regimes~\cite{raissi2019physics}, the true power of DL cannot be fully exploited by naively using the off-the-shelf DL model in the computer science community as an end-to-end fashion~\cite{hannun2014deep} for a data-fit of surrogate modeling.

In conventional ML problems, the mechanism behind the system is usually unknown and thus can only be learned from the labeled data. In contrast, for modeling a physical system, the governing equations are usually known \emph{a priori} but are difficult to solve efficiently. Instead of learning solely from the labeled data, e.g., solution of the states on certain points in parameter space, the known governing equations can be utilized to constrain (or even drive) the learning to compensate for the insufficiency of the data. Specifically, the training (optimization) of a DNN can be driven by minimizing the residual of the governing equations constructed by the DNN ansatz. This idea of physics-constrained learning is not new and was proposed back in the late '90s in the context of solving classic differential equations~\cite{lee1990neural,lagaris1998artificial,lagaris2000neural}. However, limited by the NN techniques and computational power at that time, this seminal work didn't have a big impact. Recently, this idea has been revived because of the recent advances in deep learning~\cite{lecun2015deep} combined with ever-increasing computational resources. Notably, the physics-informed neural network (PINN) proposed by Raissi et al. was used to solve a number of deterministic one-dimensional (1D) PDEs and two/three-dimensional (2D/3D) PDE-constrained inverse problems with a moderate amount of labeled data~\cite{raissi2019physics,raissi2018hidden,raissi2019deep}, e.g., noisy measurements of the velocity field. A similar approach is also applied to learn the constitutive relationship in a Darcy flow~\cite{tartakovsky2018learning}. The PINN approach has been recently extended to assimilate multi-fidelity training data~\cite{meng2019composite}, and its UQ analyses have been explored based on arbitrary polynomial chaos~\cite{zhang2018quantifying} and adversarial inference~\cite{yang2018adversarial}. Similar ideas of using physical constraints to regularize the DNN training have also been investigated in~\cite{sharma2018weakly,nabian2018physics,xu2019neural,holland2019towards}. In the aforementioned works, a moderate amount of labeled data either from simulations or experimental measurements are still needed for obtaining an approximation to the solution of the PDEs. In fact, if the initial and boundary conditions are well imposed thus the corresponding PDE problem is well-defined, in principle, the unique solution should be captured by the DNN via PDE-constrained learning without any labeled data. This is one of the main motivations of current work. Note that recently there have been several works on the concept of data-free DNNs, e.g., for solving a handful of computer vision problems~\cite{stewart2017label}, deterministic PDEs~\cite{sirignano2018dgm,berg2018unified,raissi2019physics}, high-dimensional stochastic partial differential equations (SPDE), and backward stochastic differential equations (BSDE)~\cite{weinan2017deep,beck2017machine,beck2018solving,weinan2018deep,han2018solving}.

In the context of surrogate modeling, Nabian and Meidani~\cite{nabian2018deep} and Karumuri et al.~\cite{karumuri2019simulator} applied the PDE-constrained fully-connected neural network (FC-NN) for uncertainty propagation in steady heat equations. Zhu et al.~\cite{zhu2019physics} proposed a PDE-constrained, label-free DNN surrogate model for UQ in an elliptic PDE using both the FC-NN and convolutional neural networks (CNN). Moreover, both the deterministic and probabilistic formulations of physics-constrained learning are devised and studied. Their results have shown a significant potential of using the physics-constrained DNN for surrogate modeling, where no labeled data are required during the training. Nonetheless, the success has only been demonstrated in a few model problems governed by \emph{linear elliptic} PDEs, e.g., diffusion equations. Thus it remains unclear if the physics-constrained learning can be used to handle realistic fluid systems governed by the Navier--Stokes equations in a parametric setting. Significant work is still needed to further explore the real-world problems for broad impacts.

The \emph{objective} of this paper is to develop a physics-constrained, data-free DNN for surrogate modeling of incompressible flows. A structured FC-NN is devised to approximate the solutions of the parametric Navier--Stokes equations, where the initial/boundary conditions are enforced instead of being penalized together during training in previous works~\cite{raissi2019physics}. In addition, contrary to the previous data-driven surrogates, the training of our DNN is solely driven by minimizing the residuals of the governing PDEs (i.e., conservation laws), where \emph{no expensive CFD simulation data} is used. To the best of our knowledge, the current work is the first attempt to build a parametric DNN surrogate model for fluid simulations \emph{without using any simulation data}. The effectiveness and merits of the proposed method are demonstrated by investigating a number of internal flows relevant to cardiovascular applications. The rest of the paper is organized as follows. The framework of structured FC-NN surrogate based on the physics-constrained label-free training is introduced in Section~\ref{sec:meth}. Numerical results of surrogate modeling and uncertainty propagation on several vascular flows are presented in Section~\ref{sec:result}. The performance of soft and hard boundary enforcement approaches, different adaptive activation functions, and data-free/data-driven learning strategies are discussed in Section~\ref{sec:discussion}. Finally, conclusion is drawn in Section~\ref{sec:conclusion}.

\section{Methodology}
\label{sec:meth}
\subsection{Overview}
Most low-speed flows, e.g., blood flows in large or medium sized vessels, can be described by the incompressible Navier--Stokes equations given as: 
\begin{equation}
	\label{eq:ns}
	\mathscr{F}(\mathbf{u}, p) = 0 := \left \{
	\begin{aligned}
	&\nabla \cdot \mathbf{u} = 0,  &\mathbf{x}, t \in \Omega_{f,t}, \boldsymbol{\theta} \in \mathbb{R}^d,\\
	&\frac{\partial\mathbf{u}}{\partial t} + (\mathbf{u}\cdot\nabla)\mathbf{u} + \frac{1}{\rho}\nabla p - \nu\nabla^2\mathbf{u} + \mathbf{b}_f = 0, &\mathbf{x}, t \in  \Omega_{f,t}, \boldsymbol{\theta} \in \mathbb{R}^d 
	\end{aligned} \right .
\end{equation}
where $t$ and $\mathbf{x}$ are time and space coordinates, respectively; $\Omega_{f,t} \triangleq \Omega_f \times [0, T]$; $\boldsymbol{\theta}$ is a $d$-dimensional parameter vector, including input and/or operational parameters such as fluid properties, inlets/outlets, and geometry of the domain; both velocity $\mathbf{u}(t, \mathbf{x},\boldsymbol{\theta})$ and pressure $p(t, \mathbf{x}, \boldsymbol{\theta})$ are functions of time, space, variable parameters; $\rho$ and $\nu$ represent density and viscosity of the fluid, respectively; $\mathbf{b}_f$ is the body force; $\Omega_f \subset \mathbb{R}^3$ denotes the fluid domain. The solutions of velocity $\mathbf{u}$ and pressure $p$ can be uniquely determined when suitable initial and boundary conditions are prescribed,
\begin{subequations}
    \label{eq:bc}
	\begin{alignat}{2}
	\mathcal{I}(\mathbf{x}, p, \mathbf{u}, \boldsymbol{\theta}) &= 0, \qquad & &\mathbf{x} \in \Omega_f, t =0,\boldsymbol{\theta} \in \mathbb{R}^d,\\
	\mathcal{B}(t, \mathbf{x}, p, \mathbf{u}, \boldsymbol{\theta}) &= 0, \qquad & &\mathbf{x}, t \in \partial\Omega_f \times [0, T], \boldsymbol{\theta} \in \mathbb{R}^d,
	\end{alignat}
\end{subequations}
where both $\mathcal{I}$ and $\mathcal{B}$ are general differential operators that define the initial and boundary conditions, respectively; $\partial\Omega_f$ denotes the boundary region. When a set of parameters $\boldsymbol{\theta}$ is given, the flow field, i.e., $\mathbf{u}(t, \mathbf{x})$ and $p(t, \mathbf{x})$, can be solved numerically by discretizing the Eqs.~\ref{eq:ns} and~\ref{eq:bc} using FD/FV/FE methods. However, this process involves mesh generation and iteratively solving large linear/nonlinear systems, which is usually time-consuming. Therefore, propagating the parameter uncertainty or inferring the unknown parameters through the FD/FV/FE solver becomes intractable when it comes to parametric problems, e.g., some parameters of $\boldsymbol{\theta}$ are uncertain or unknown.  Solving varying-geometry problems is especially challenging since any change of the geometry requires regeneration of the computational meshes. 
\begin{figure}[htbp]
	\centering
	\includegraphics[width=0.7\textwidth]{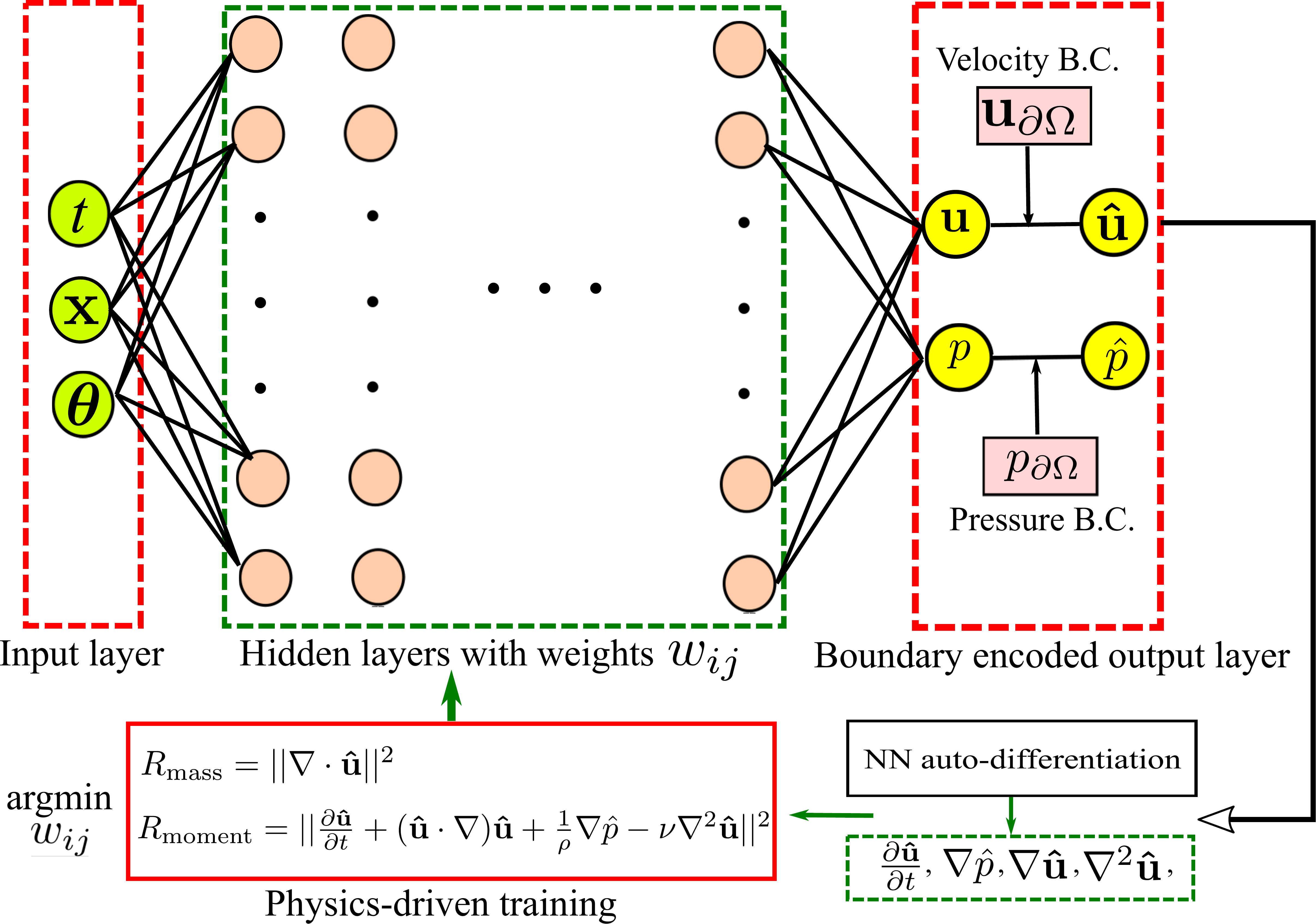}
	\caption{A schematic diagram of the physics-constrained, data-free DL framework for surrogate modeling of fluid flows. A structured fully-connected neural network (FC-NN) is developed with the boundary conditions encoded by construction. The network is trained by minimizing the equation-based loss function and no CFD simulation data are needed.}
	\label{fig:frame}
\end{figure}
To enable fast predictions in terms of UQ and optimization applications, a deep neural network (DNN) architecture is built to approximate the solutions of the Navier--Stokes equations in a parametric setting. The DNN-based surrogate is expected to provide a rapid online prediction of the flow field with any given set of parameters $\boldsymbol{\theta}$ after the offline training. A schematic diagram of the proposed framework is shown in Fig.~\ref{fig:frame}. A FC-NN is devised with the input layer composed of time $t$, spatial coordinates $\mathbf{x}$, and variable parameters $\boldsymbol{\theta}$. The raw outputs of the FC-NN (i.e., $\mathbf{u}$ and $p$) are used to construct the state variables (i.e., velocity $\mathbf{\hat{u}}$ and pressure $\hat{p}$) together with contribution from the particular solution that encodes initial/boundary conditions. The FC-NN is trained by minimizing residuals of the Navier--Stokes equations and no data from CFD simulations are needed. Therefore, the DNN predictions are expected to conform to the conservation laws of fluid flows and satisfy the specified initial/boundary conditions. Note that the Navier--Stokes equations will not be solved with any numerical discretization. The details of the physics-constrained training and boundary condition enforcement will be presented in the following subsections.    

\subsection{Deep Neural Network and Physics-Constrained Training}
Neural networks (NN) are a set of algorithms, inspired by the biological neural networks in brains, for classification and regression tasks. There are various types of NNs with different neuron connection forms and architectures, e.g., fully-connected neural networks (FC-NN), convolutional neural networks (CNN), and recurrent neural networks (RNN). In this work, the feedforward FC-NN is considered, where the neurons of adjacent layers are fully connected and outputs of each layer are fed forward as the inputs to the next layer. A FC-NN defines a mapping from the input layer $\mathbf{z}_0 \in \mathbb{R}^{n_0}$ to the output $\mathbf{z}_{L} \in \mathbb{R}^{n_L}$. The layers between the input and output layers are called hidden layers $\mathbf{z}_l$, where $l = 1,\ldots,L-1$. By convention, a neural network with more than one hidden layer is called a ``deep" NN. Mathematically, two adjacent layers are connected as,
\begin{equation}
    \mathbf{z}_l = \sigma_l( \mathbf{W}_l^T\mathbf{z}_{l-1} + \mathbf{b}_l), 
    \label{eq:nn}
\end{equation}
where $\mathbf{W}_l \in \mathbb{R}^{n_{l-1} \times n_{l}}$ and $\mathbf{b}_l \in \mathbb{R}^{n_l}$ are the weight matrix and bias vector; the subscript $l$ denotes the index of the layer; $\sigma_l(\cdot)$ is an activation function acting element-wise, for which a number of options can be chosen, e.g., sigmoids, rectified linear units (ReLU), and tanh functions. After training, the weights, bias, and activation function at each layer are determined, and the output prediction $\mathbf{z}_L$ (i.e., velocity and pressure) can be rapidly computed from any given input vector $\mathbf{z}_0$ (i.e., coordinates and parameters) based on the Eq.~\ref{eq:nn}. Since this feedforward algorithm (Eq.~\ref{eq:nn}) only involves a few matrix multiplications, the computational cost for evaluating the trained FC-NN can be neglected compared to that of a CFD simulation.

Traditionally, to build a surrogate model for the CFD simulation of the solution $\mathbf{f}(t, \mathbf{x}, \boldsymbol{\theta})$, one can simply consider a black-box surrogate, e.g., FC-NN, or CNN~\cite{zhu2018bayesian}, as  $\mathbf{z}_L(t, \mathbf{x}, \boldsymbol{\theta}; \mathbf{W}, \mathbf{b})$, i.e., 
\begin{equation}
    \mathbf{f}(t, \mathbf{x}, \boldsymbol{\theta}) \approx \mathbf{f}^p(t, \mathbf{x}, \boldsymbol{\theta}) \triangleq \mathbf{z}_L(t, \mathbf{x}, \boldsymbol{\theta}; \mathbf{W}, \mathbf{b}),
\end{equation}
where $\mathbf{f}$ is the solution vector as 
$\mathbf{f} = \begin{bmatrix}
\mathbf{u} &
p
\end{bmatrix}^\intercal$, including velocity $\mathbf{u}$ and pressure $p$; $\mathbf{W}$ and $\mathbf{b}$ denote the weights and biases of the entire network. Generally, training of a DNN is purely data-driven, and it consists of finding a set of (sub)optimal DNN parameters ($\mathbf{W}, \mathbf{b}$) such that the mismatch between the training data $\mathbf{f}^d$ and the DNN predictions $\mathbf{f}^{p}$ is locally minimized. That is, one can formulate an optimization problem as,
\begin{subequations}
\label{eq:op1}
  \begin{alignat}{2}
    \mathcal{L}_{data}(\mathbf{W}, \mathbf{b}) &= \norm{\mathbf{f}^d(t, \mathbf{x}, \boldsymbol{\theta}) - \mathbf{z}_L(t, \mathbf{x}, \boldsymbol{\theta}; \mathbf{W}, \mathbf{b})}_{\Omega_{f,t}}, \\
    \mathbf{W}^*, \mathbf{b}^* &= \underset{\mathbf{W}, \mathbf{b}}{\arg\min} \ \mathcal{L}_{data}(\mathbf{W}, \mathbf{b}),
  \end{alignat}
\end{subequations}
where the loss function $\mathcal{L}_{data}(\mathbf{W}, \mathbf{b})$ is named as ``data-based loss", and $\norm{\cdot}_{\Omega_{f,t}}$ is L2 norm over $\Omega_{f,t}$; $\mathbf{W}^*, \mathbf{b}^*$ denote a set of (sub)optimal NN weights and biases obtained from the optimization. 

However, as discussed above, this black-box surrogate modeling requires enormous training data $\mathbf{f}^d$, which is too expensive to obtain from a large number of CFD simulations. Instead, following previous PINN framework~\cite{raissi2019physics}, we consider leveraging the governing PDEs in the loss function by minimizing the violation of the solution $\mathbf{z}_L$ in terms of the known governing PDEs for fluid dynamics over a domain of interests without the needs of solving these equations for each parameter with traditional numerical methods. Specifically, only the residuals of the Navier--Stokes equations are computed based on the FC-NN predictions and it corresponds to a constrained optimization as follows,
\begin{equation}
\label{eq:op2}
  \begin{aligned}
    \mathcal{L}_{phy}(\mathbf{W}, \mathbf{b}) &= \underbrace{\norm{\nabla \cdot \mathbf{u}^p}_{\Omega_{f,t}}}_{\mathrm{Mass \ conservation}} + \underbrace{\lVert\frac{\partial\mathbf{u}^p}{\partial t} + (\mathbf{u}^p\cdot\nabla)\mathbf{u}^p + \frac{1}{\rho}\nabla p^p - \nu\nabla^2\mathbf{u}^p + \mathbf{b}_f\rVert_{\Omega_{f,t}}}_{\mathrm{Momentum \ conservation}},
  \end{aligned}
\end{equation}
\begin{equation}
\label{eq:op2-c}
	\begin{aligned}
	&\mathbf{W}^* = \underset{\mathbf{W}}{\arg\min} \ \mathcal{L}_{phy}(\mathbf{W}),\\
		&\mathrm{s. t.}
		\left \{
		\begin{split}
		&\mathcal{I}(\mathbf{x}, p^p, \mathbf{u}^p, \boldsymbol{\theta}) = 0,  \qquad &t =0, \mathrm{in}\ \Omega_{f},\\
		&\mathcal{B}(t, \mathbf{x}, p^p, \mathbf{u}^p, \boldsymbol{\theta}) = 0, &\mathrm{on}\ \partial\Omega_{f,t},
		\end{split} \right.
	\end{aligned}
\end{equation}
where the loss function $\mathcal{L}_{phy}(\mathbf{W}, \mathbf{b})$ here is named as ``physics-based loss"; the superscript $p$ indicates that the quantities are predicted by the DNN. To construct the PDE residuals in the loss function, several first and/or second derivative terms of $\mathbf{u}^p$ and $p^p$ with respect to time $t$ and space coordinates  $\mathbf{x}$ are required, which can be computed based on automatic differentiation (AD)~\cite{baydin2018automatic}. AD is an accurate and efficient way to calculate derivatives in a computational graph, which has started to gain increasing attention in the machine learning community. The general idea of AD is to use the chain rule to back-propagate derivatives from the output layer to the inputs as the connection between each layer of a NN is analytically defined. Compared to numerical differentiation techniques, derivatives calculated from AD are much more accurate since they do not suffer from truncation or round-off errors. Most modern deep learning frameworks such as PyTorch~\cite{paszke2017automatic}, TensorFlow~\cite{abadi2016tensorflow}, and Theano~\cite{bastien2012theano} have the AD implemented. To solve the optimization problem defined in Eq.~\ref{eq:op2}, stochastic gradient descent (SGD) algorithms are used, which are known to be a stochastic approximation of the gradient descent (GD) optimization. In SGD, only a subset of points are randomly sampled from the input space to calculate the direction of the gradient at each iteration. The SGD algorithms are known to work very well to escape bad local minima in the neural network training~\cite{kleinberg2018alternative} under one point convexity property. Although the global minimum cannot be guaranteed for a non-convex optimization problem as defined in Eq.~\ref{eq:op2}, an empirically good local minimum is usually found based on the SGD algorithms.

\subsection{Boundary Condition Enforcement}
If the physics-based loss $\mathcal{L}_{phy}$ becomes identically zero, the DNN predictions of velocity $\mathbf{u}^p$ and pressure $p^p$ will exactly satisfy the Navier--Stokes equations (Eq.~\ref{eq:ns}). Therefore, penalizing the PDE residuals can regularize the data-driven DNN solutions to be more physical. This idea is known as the physics-informed, weakly-supervised deep learning~\cite{raissi2019physics,raissi2018hidden,nabian2018physics}. 
To make the problem well-posed, proper initial and boundary conditions (IC/BC) are needed and imposed as constraints (Eq.~\ref{eq:op2-c}) which are often treated in a ``soft" manner by modifying the original loss function with penalty terms~\cite{marquez2017imposing,raissi2019physics}. For example, the IC/BC can be imposed in a ``soft" way by modifying Eq.~\ref{eq:op2} as,     
\begin{equation}
\label{eq:op25}
    \mathcal{L}_{phy}^c(\mathbf{W}, \mathbf{b}, \lambda_i, \lambda_b) = \underbrace{\mathcal{L}_{phy}(\mathbf{W}, \mathbf{b})}_{\mathrm{Equation \ loss}} + \underbrace{\lambda_i \norm{\mathcal{I}(\mathbf{x}, p^p, \mathbf{u}^p, \boldsymbol{\theta})}_{\Omega_{f,t}}}_{\mathrm{Initial \ loss}} + \underbrace{\lambda_b\norm{\mathcal{B}(t, \mathbf{x}, p^p, \mathbf{u}^p, \boldsymbol{\theta})}_{\partial\Omega_{f,t}}}_{\mathrm{Boundary \ loss}},
\end{equation}
where $\lambda_i$ and $\lambda_b$ are penalty coefficients. However, the soft IC/BC enforcement methods have several major drawbacks: (1) there is no quantitative guarantee on how accurate the IC/BC being imposed and thus the solution could be unsatisfactory; (2) the optimization performance can depend on the relative importance of each term, but how to assign weight for each term can be difficult. Alternatively, we can impose the IC/BC in a ``hard" manner, where a particular solution that solely satisfies the initial/boundary condition is added. Hence, the constraints on IC/BC are automatically fulfilled. A mixed enforcement on IC/BC is proposed in this work, where the Neumann and Dirichlet boundary conditions (BC) are treated separately: the Neumann BC are formulated into the equation loss, i.e., in a soft manner, while the IC and Dirichlet BC are encoded in a hard manner by constructing the DNN ansatz $\hat{\mathbf{u}}^p$ and $\hat{p}^p$ with a particular solution as follows,
\begin{align}
\label{eq:nn_hard}
 \begin{split}
    \hat{\mathbf{u}}^p(t, \mathbf{x}, \boldsymbol{\theta};\mathbf{W}, \mathbf{b}) &= \mathbf{u}_{particular}(t, \mathbf{x}, \mathbf{\theta}) + D(t, \mathbf{x}, \boldsymbol{\theta})\mathbf{u}^p(t, \mathbf{x}, \boldsymbol{\theta};\mathbf{W}, \mathbf{b}),\\
    \hat{p}^p(t, \mathbf{x}, \boldsymbol{\theta};\mathbf{W}, \mathbf{b}) &= p_{particular}(t, \mathbf{x}, \mathbf{\theta}) +  D(t, \mathbf{x}, \boldsymbol{\theta}){p}^p(t, \mathbf{x}, \boldsymbol{\theta};\mathbf{W}, \mathbf{b}) ,
 \end{split}
\end{align}
where $\mathbf{u}_{particular}$ is a particular solution that just satisfies IC and BC: $\mathbf{u}_{particular}(\mathbf{x}, 0) = \mathbf{u}_0(\mathbf{x})$,  $\mathbf{p}_{particular}(\mathbf{x}, 0) = p_0(\mathbf{x})$ and $\mathbf{u}_{particular}(\mathbf{x}, t)\vert_{\mathbf{x} \in \partial \Omega_f} = \mathbf{u}^b(\mathbf{x})$,  $\mathbf{p}_{particular}(\mathbf{x}, t)\vert_{\mathbf{x} \in \partial \Omega_f} = p^b(\mathbf{x})$; $D(t, \mathbf{x}, \boldsymbol{\theta})$ is a globally defined smooth function from internal points to the ``boundary" in $\Omega_{f,t}$, i.e., a space--time sense. That is, $D$ is zero on the boundary $\partial \Omega_f \times [0, T]$ and $ \Omega_f \times \{0\}$ while increases away from the boundary. For those problems where the IC/BC and the geometry of the domain in $\Omega_{f,t}$ is simple, the function $D$ and particular solution can be written analytically. However, if the geometry is too complex to have an analytic form, e.g., a patient-specific artery, the function $D$ can be pre-trained by a low-capacity NN in the way proposed in~\cite{berg2018unified}. Finally, the constrained optimization problem in Eq.~\ref{eq:op2} can be reformulated as an unconstrained one, as shown in Eq.~\ref{eq:op3}.
\begin{subequations}
\label{eq:op3}
  \begin{alignat}{2}
    \begin{split}
    \mathcal{L}^c_{phy}(\mathbf{W}, \mathbf{b}) = &\underbrace{\norm{\nabla \cdot \hat{\mathbf{u}}^p}_{\Omega_f}}_{\mathrm{Mass \ conservation}} + \underbrace{\lVert\frac{\partial\hat{\mathbf{u}}^p}{\partial t} + (\hat{\mathbf{u}}^p\cdot\nabla)\hat{\mathbf{u}}^p + \frac{1}{\rho}\nabla \hat{p}^p - \nu\nabla^2\hat{\mathbf{u}}^p + \mathbf{b}\rVert_{\Omega_f}}_{\mathrm{Momentum \ conservation}} \\
    &+  \underbrace{\norm{\mathcal{B}_N(t, \mathbf{x}, \hat{p}^p, \hat{\mathbf{u}}^p, \boldsymbol{\theta})}_{\partial\Omega_f}}_{\mathrm{Newman\ boundary \ loss}},
    \end{split}\\
    \begin{split}
    \mathbf{W}^*, \mathbf{b}^* &= \underset{\mathbf{W}, \mathbf{b}}{\arg\min} \ \mathcal{L}^c_{phy}(\mathbf{W}, \mathbf{b}).
    \end{split}
  \end{alignat}
\end{subequations}

\section{Numerical Results}
\label{sec:result}
A number of 2D vascular flows with idealized geometries, including circular pipe flows, stenotic flows, and aneurysmal flows, are studied to evaluate the performance of the proposed data-free DL surrogate model in Eq.~\ref{eq:op3}. Forward propagation of the uncertainties in the fluid properties and domain geometry to the flow QoIs (e.g., velocity, pressure, and wall shear stress) are investigated through the DNN surrogate model. In this study, only steady-state solutions are considered for proof-of-concept, thus the constraint of initial condition can be neglected in Eq.~\ref{eq:nn_hard}.

A composite FC-NN architecture is devised for the surrogate, which is composed of three sub-DNNs with an identical structure of 3 hidden layers with 20 neurons per layer. The Swish activation function~\cite{ramachandran2017swish} with fixed hyperparameters is employed in each layer except the last one, where a linear activation function is used. The three sub-DNNs share the same input layer and separately predict three scalar state variables, i.e., velocity $u,v$, and pressure $p$. All three sub-DNNs are trained simultaneously with a unified physics-based loss function. To solve the unconstrained optimization problem defined in Eq.~\ref{eq:op3}, we used the Adam optimizer~\cite{kingma2014adam}, a robust variant of the SGD method, where the learning rate is adaptively changed based on the estimates of the moments. The initial learning rate and mini-batch size is set as $1\times10^{-3}$ and 50, respectively. Because of the adaptivity feature of the Adam optimizer, the hyperparameters in the training are robust to some extent and require little tuning. Properly initializing the DNN parameters is also important. After comparing several widely-used initialization schemes~\cite{Glorot10understandingthe,he2015delving}, we chose the He's normal initializer~\cite{he2015delving}, where initial weights are drawn from a truncated normal distribution. In general, a good choice of the DNN architecture, including the number of layers, number of neurons per layer, activation function, and initialization schemes, is important to the learning performance but is still determined by trial and error. A rule of thumb is to achieve a favorable performance using the simplest network structure, which enables fast training and better generalizability. In this work, a handful of DNN architectures with different ``depths" are investigated, and the ``shallowest" one with a satisfactory learning performance is adopted. To demonstrate the robustness of the physics-constrained learning, the architecture and hyperparameters remain the same for all the cases throughout the paper. Note that a comprehensive parameter study and architecture optimization of the DNN is out of the scope of the current work.

The composite FC-NN is implemented in the PyTorch platform~\cite{paszke2017automatic}. As discussed in Section~\ref{sec:meth}, only the collocation points are required and they are uniformly sampled in the spatial $\mathbf{x}$ and parameter $\boldsymbol{\theta}$ spaces. Alternatively, one can choose a space-filling Latin hypercube sampling in the $\Omega_{f}$~\cite{raissi2019physics}. In this work, the PDE residuals are extensively evaluated on a large number of collocation points to ensure learning quality. For all test cases, the training of about $10^6$ SGD iterations are performed on an NVIDIA GeForce GTX 1080Ti Graphics Processing Unit (GPU) card, and the cost is approximately 3.5 hours. Note that the offline training cost can be potentially reduced by optimizing the DNN architecture. To validate the prediction performance of the trained DNN surrogates, corresponding CFD simulations are also conducted using an open-source FV-based CFD solver, OpenFOAM~\cite{Jasak07openfoam:a}. Mesh convergence study is performed to ensure the solution accuracy. The code and datasets for this work will become available at \texttt{https://github.com/Jianxun-Wang/LabelFree-DNN-Surrogate} upon publication.          

\subsection{Circular Pipe Flow}
\label{sec:Circular Pipe}
Flow in a pipe/tube is very common in physiological systems, e.g., blood in arteries or air flow in trachea. The pipe flow is often driven by the pressure difference between the two ends of a tube, or by the body force of gravity. In a cardiovascular system, the former one is more dominant since the blood flow is mainly governed by pressure drop due to the heart pumping. In general, simulating the fluid dynamics in a tube requires solving the full Navier--Stokes equations numerically, but if the tube is straight and has a constant circular cross section, analytical solution of the fully-developed steady-state flow is available, which is an ideal benchmark to validate the performance of the proposed method. As a result, we first study the flow in a 2D circular pipe (also known as the Poiseuille flow).    

In this case, the pressure inlet and outlet are used to drive the flow since we only focus on the fully-developed regime, and no-slip wall boundary is prescribed on the tube walls. The boundary conditions are encoded into the surrogate model by constructing the DNN ansatz $\hat{u}^p, \hat{v}^p$, and $\hat{p}^p$ based on Eq.~\ref{eq:nn_hard}. The no-slip condition of velocity on the wall can be imposed by designing the $\hat{u}^p, \hat{v}^p$ as, 
\begin{equation}
\label{eq:u_hard}
\hat{u}^p = \left(\frac{d^2}{4} - y^2\right) u^p,  \qquad \hat{v}^p = \left(\frac{d^2}{4} - y^2\right) v^p,
\end{equation}
where $y$ is the radial distance, $d = 0.1$ is the diameter of the tube,  the raw DNN output is denoted by $u^p$.
The pressure inlet $p_{in} = 0.1$ and outlet $p_{out} = 0$ are imposed by designing the $\hat{p}^p$ as,
\begin{equation}
\label{eq:p_hard}
    \hat{p}^p = \frac{x - x_{in}}{x_{out} - x_{in}} p_{out} + \frac{x_{out} - x}{x_{out}- x_{in}} p_{in} + (x - x_{in})(x_{out} - x) p^p,
\end{equation}
where $x_{in}$ and $x_{out}$ are coordinates of the two ends of the tube, and the raw DNN output is denoted by $p^p$. All three sub-DNNs are trained to capture the spatial flow fields with parameter variation in the fluid viscosity $\nu$. Input (collocation) points in the parameter space of $\nu$ are uniformly sampled in the range $10^{-4}\le \nu \le 1.9\times10^{-3}$, where the corresponding Reynolds number (Re) are moderate ({Re $< 300$}). After training, both velocity and pressure fields can be obtained immediately by evaluating the trained DNN with any given input $\nu$ and a spatial grid on $\mathbf{x}$. Hence, the DNN surrogate can be utilized to propagate the uncertainty in viscosity $\nu$ based on the Monte Carlo (MC) simulation, where a large number of samples are drawn from the $\nu$ distribution and propagated to the QoIs via the DNN surrogate. In the following test cases, 500 MC samples are used to compute desired statistics and distributions. 
\begin{figure}[htbp]
  \centering
    \subfloat[Cross-section velocity profiles]{\includegraphics[width=0.4\textwidth]{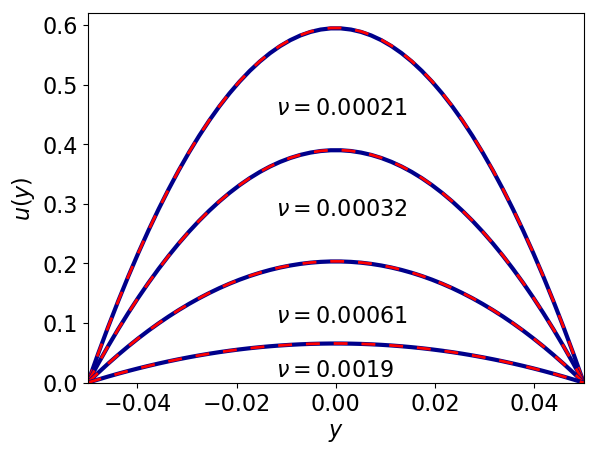}}
    \subfloat[Distribution of center velocity] {\includegraphics[width=0.4\textwidth]{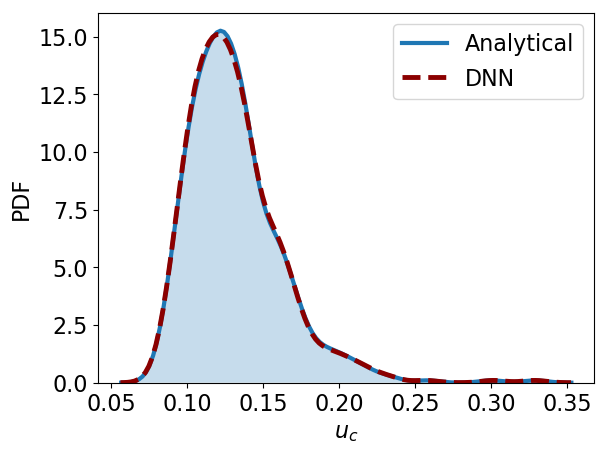}}
    \caption{(a) DNN predicted cross-section pipe flow velocity profiles $u(y)$ of four different viscosity ($\nu$) samples compared with the analytical solution; (b) probability density function of the center velocity $u_c = u(y=0)$ propagated from a normally distributed $\nu$ with mean $\overline{\nu} = 10^{-3}$ and variance of $\sigma_{\nu} = 2\times10^{-4}$, using the trained DNN surrogate compared with the analytical solution.}
  \label{fig:pipe}
\end{figure}
The DNN surrogate results (shown in Figure~\ref{fig:pipe}) are compared against the analytical solution, which is given by
\begin{equation}
\label{eq:pipe_analytical}
    u_a = \frac{\Delta p}{2 \nu \rho L}\left(\frac{d^2}{4} - y^2\right),
\end{equation}
where $y$ denotes the spanwise coordinate and $\Delta p$ is the pressure difference. It can be observed from Fig.~\ref{fig:pipe}a that the DNN-predicted velocity profiles (red dashed lines) of four different $\nu$ samples almost exactly agree with the analytical solutions (blue solid lines), where the Reynolds numbers (Re) of the four cases are 283, 121, 33, and 3, respectively. Actually, the trained DNN is able to accurately predict the pipe flow field with any given viscosity, where the Reynolds number is moderate. Fig.~\ref{fig:pipe}b shows the uncertainty of the center velocity $u_c$ propagated from a normally distributed $\nu$ with mean of $\overline{\nu} = 1\times10^{-3}$ and variance of $\sigma_{\nu} = 2\times10^{-4}$. It can be seen that the DNN-predicted probability density function (PDF) is almost identical to that of the analytical solutions, demonstrating the excellent performance of the physics-constrained learning for uncertainty propagation. 

\subsection{Blood Flow with Standardized Geometry}
\label{sec:3.2}
Two types of canonical vascular flows, stenotic flow and aneurysmal flow, with standardized vessel geometries are studied. A stenotic flow refers to the flow through a vessel where there is narrowing and re-expansion of the vessel wall. This local restriction of the vessel is related to many cardiovascular diseases, e.g., arteriosclerosis, stroke, and heart attack~\cite{berger2000flows}. The vascular flow within an aneurysm, which is the expansion of an artery due to the weakness of vessel walls, is called aneurysmal flow. The rupture of an aneurysm may cause life-threatening conditions, e.g., subarachnoid hemorrhage (SAH) due to cerebral aneurysm rupture~\cite{brisman2006cerebral}, and investigation of the hemodynamics can improve the diagnosis and fundamental understanding of aneurysm progression and rupture~\cite{cebral2011association}.  

Whereas realistic vascular geometries are usually irregular and complex, including sites of curvature, bifurcation and junctions, idealized stenosis and aneurysm models are studied here for proof-of-concept. Namely, both the stenotic and aneurysmal vessels are idealized as an axisymmetric tube with a varying cross-section radius, which is parameterized by the following function,
\begin{equation}
\label{eq:radius_ste&equ}
    R(x) = R_0 - A\frac{1}{\sqrt{(2\pi\sigma^2)}}\exp(-\frac{(x-\mu)^2}{2\sigma^2})
\end{equation}
where $R_0$ is the radius at the inlet, which is set as 0.05, and the sign of $A$ determines if the vessel is stenotic or aneurysmal. Namely, a positive and negative sign correspond to the stenosis and aneurysm, respectively. Three control parameters $A$, $\mu$, and $\sigma$ define the shape of the stenotic (aneurysmal) vessel. The scale parameter $A$ controls the curvature along the tube, and a larger $|A|$ leads to narrower stenosis (broader aneurysm). Parameter $\mu$ defines the streamwise location of the minimum (maximum) radius of the stenosis (aneurysm), and $\sigma$ affects the steepness of the geometric variation. In this study, the latter two parameters $\mu$ and $\sigma$ are fixed as $0.5$ and $0.1$, respectively. Only $A$ is considered as a variable parameter to control the degree of the stenosis (aneurysm).

Similar to the pipe flow, the pressure inlet/outlet and no-slip wall boundary conditions are prescribed for both stenosis and aneurysm cases. The boundary-encoded sub-DNNs are constructed to learn the parametric flow solutions, where the wall BC is imposed using the geometric function $D(x, y) = R(x)^2 - y^2$. Contrary to the case studied above, where only the viscosity variation is considered, the DNN surrogate is also trained to capture the varying stenosis (aneurysm) geometry, which is known challenging for mesh-based CFD simulations. Specifically, the solutions of varying viscosity are learned for fixed vascular geometries ($A = 5\times10^{-3}$ for stenosis and $A = -5\times10^{-3}$ for aneurysm) in the first place, and then the performance for capturing geometry variations is examined at a fixed viscosity ($\nu = 1\times10^{-3}$). Moreover, the uncertainties from the flow viscosity and vessel geometry are propagated to the QoIs through the trained DNN surrogate using MC sampling, and the results are validated by CFD-based MC simulations. 

\subsubsection{Flow in Idealized Stenosis}
\label{sec:3.2.1}
The DNN is trained to parameterize the solutions of stenotic flows with varying viscosity, where collocation points are sampled in $\nu$ space within the range of $[5\times10^{-4}, 1\times10^{-2}]$ for physics-based training.
\floatsetup[figure]{style=plain,subcapbesideposition=top}
\fboxsep= 0pt
\fboxrule=0.1pt
\begin{figure}[htb]
	\centering 
	\sidesubfloat[]{\fbox{\includegraphics[width=0.28\textwidth]{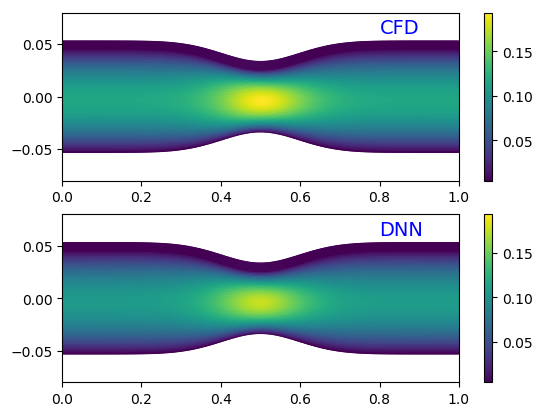}
	 \includegraphics[width=0.28\textwidth]{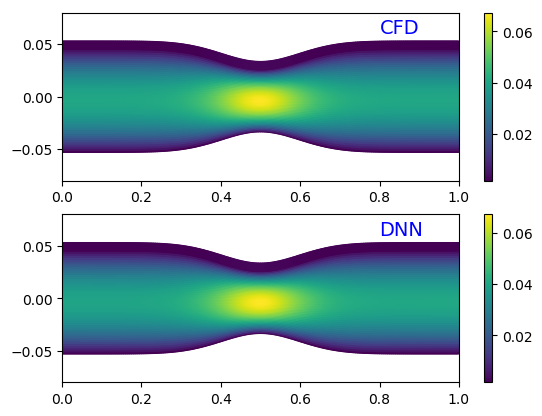}
	\includegraphics[width=0.28\textwidth]{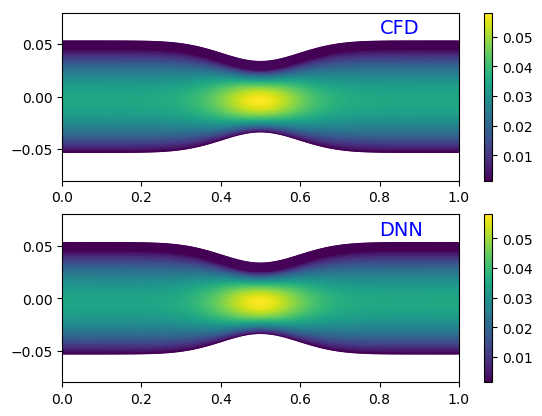}}}
	\par\medskip 
	\vfill
	\sidesubfloat[]{\fbox{\includegraphics[width=0.28\textwidth]{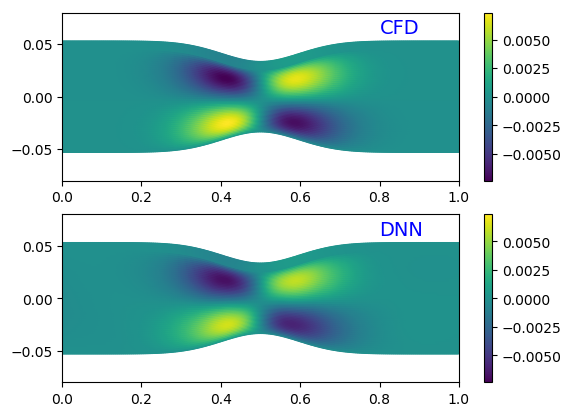}
	\includegraphics[width=0.28\textwidth]{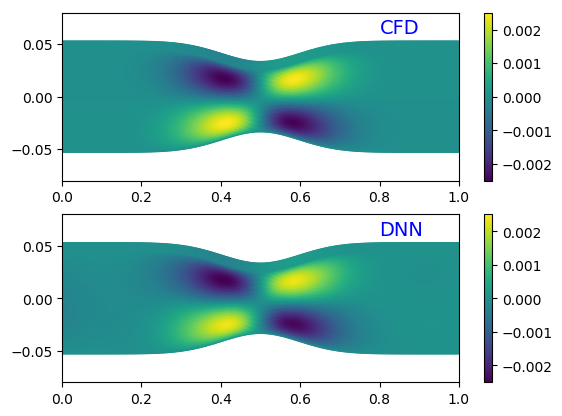}
	\includegraphics[width=0.28\textwidth]{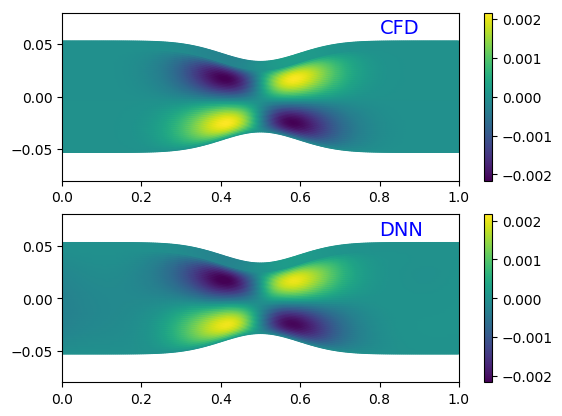}}}
	\par\medskip 
	\vfill
	\sidesubfloat[]{\fbox{\includegraphics[width=0.28\textwidth]{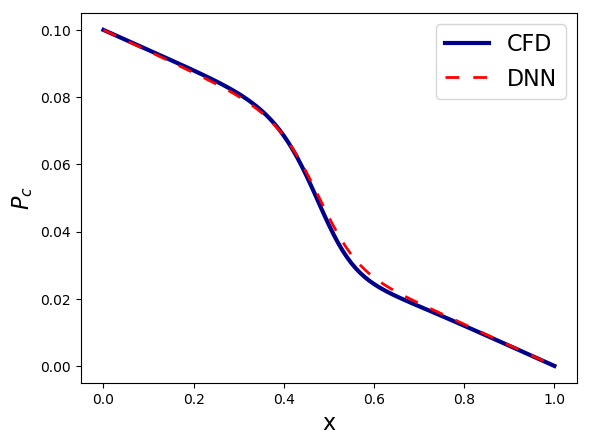}
	\includegraphics[width=0.28\textwidth]{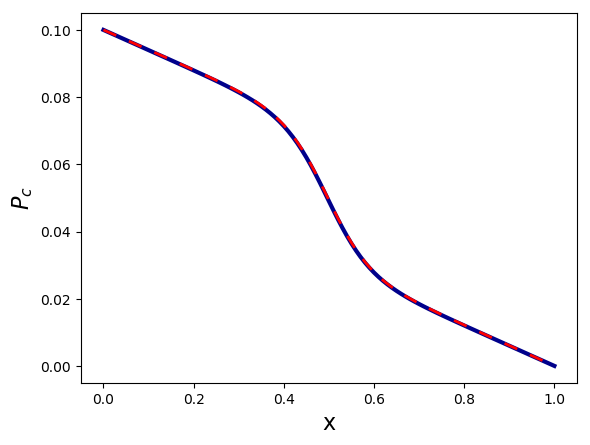}
	\includegraphics[width=0.28\textwidth]{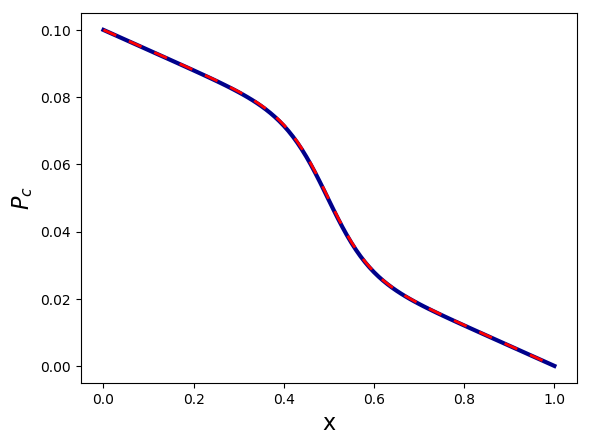}}}
	\caption{Comparison between physics-constrained DNN predictions and CFD solutions of idealized stenotic flows at three different viscosity samples: (left) $\nu = 6.4\times10^{-4}$, (middle) $\nu = 1.85\times10^{-3}$, (right) $\nu = 2.14\times10^{-3}$. (a) streamwise velocity component $u$ (b) spanwise velocity component $v$ (c) centerline pressure profile $P_c$.}
	\label{fig:stenotic_vis}
\end{figure}
Figure~\ref{fig:stenotic_vis} shows the DNN-predicted flow fields of three different viscosity samples, i.e., $\nu = 6.4\times10^{-4}$, $1.85\times10^{-3}$, and $2.14\times10^{-3}$, at moderate Reynolds numbers. The corresponding CFD simulations are performed for comparison. It can be seen that the flow patterns of different $\nu$ are similar, where the fluid is accelerated streamwisely through the converging region and slows down passing the diverging part of the tube. However, the velocity magnitude reduces as the viscosity increases (left to right columns). As shown in Figs~\ref{fig:stenotic_vis}a and~\ref{fig:stenotic_vis}b, the DNN-predicted velocity contours of streamwise and spanwise components ($u$ and $v$) agree with the CFD solutions very well, though the magnitude in the case with the smallest viscosity ($\nu = 6.4\times10^{-4}$) is slightly underestimated. Moreover, the nonlinear pressure drops can be accurately captured by the DNN surrogate as the profiles of centerline pressure from the DNN and CFD are almost identical (Fig.~\ref{fig:stenotic_vis}c).     

 To learn the flow solutions with varying geometry, the DNN is trained on uniformly sampled points within the range of $0 \le A \le 1\times10^{-2}$. 
\begin{figure}[htb]
	\centering 
	\sidesubfloat[]{\fbox{\includegraphics[width=0.28\textwidth]{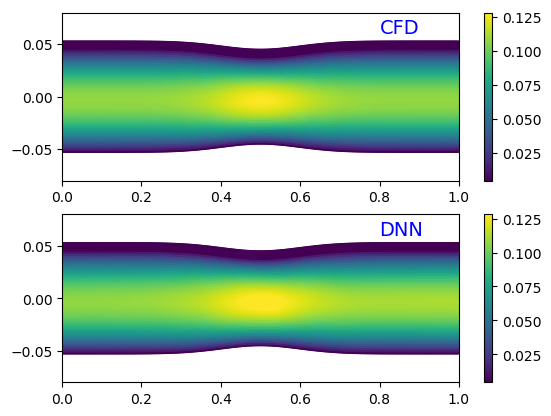}
	\includegraphics[width=0.28\textwidth]{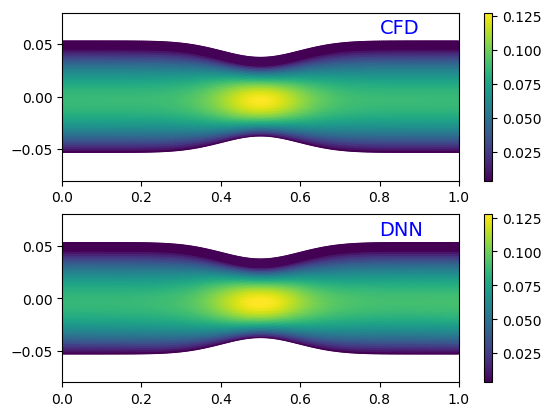}
	\includegraphics[width=0.28\textwidth]{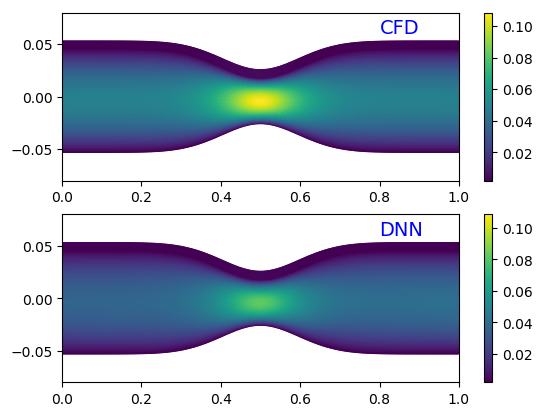}}}
	\par\medskip
	\vfill
	\sidesubfloat[]{\fbox{\includegraphics[width=0.28\textwidth]{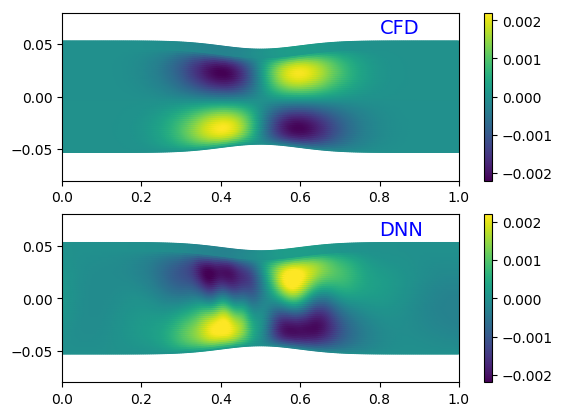}
	\includegraphics[width=0.28\textwidth]{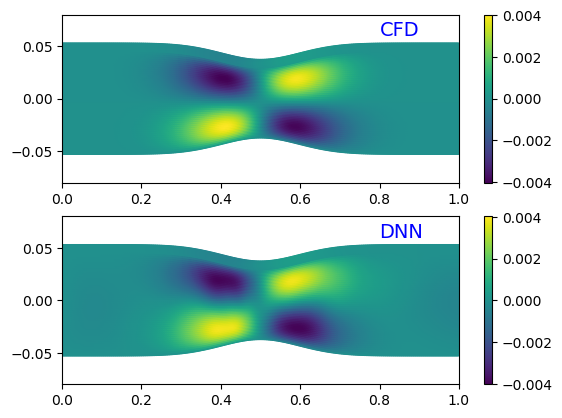}
	\includegraphics[width=0.28\textwidth]{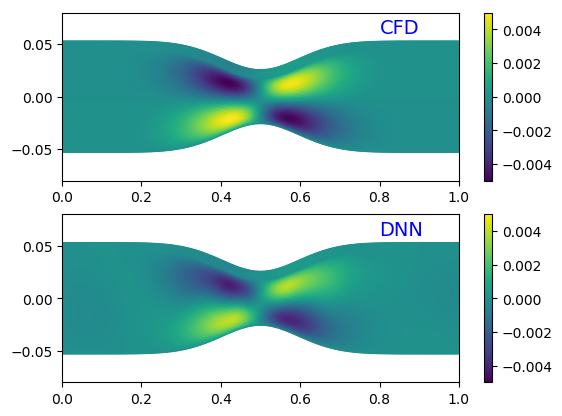}}}
	\par\medskip
	\vfill
	\sidesubfloat[]{\fbox{\includegraphics[width=0.28\textwidth]{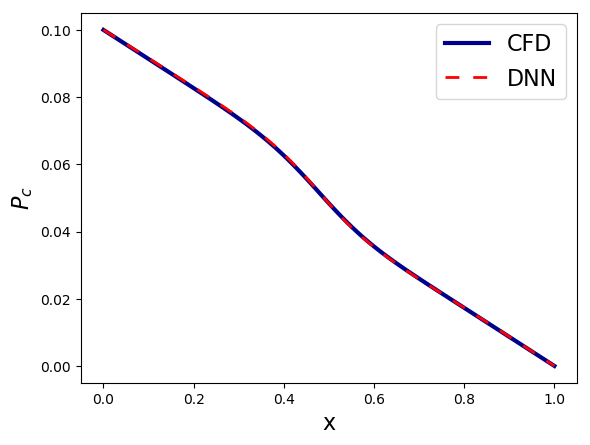}
	\includegraphics[width=0.28\textwidth]{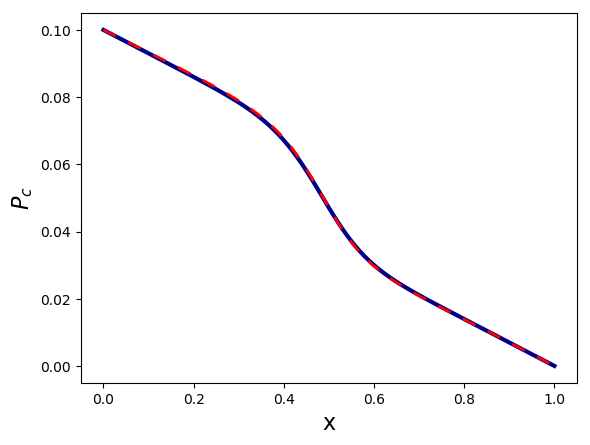}
	\includegraphics[width=0.28\textwidth]{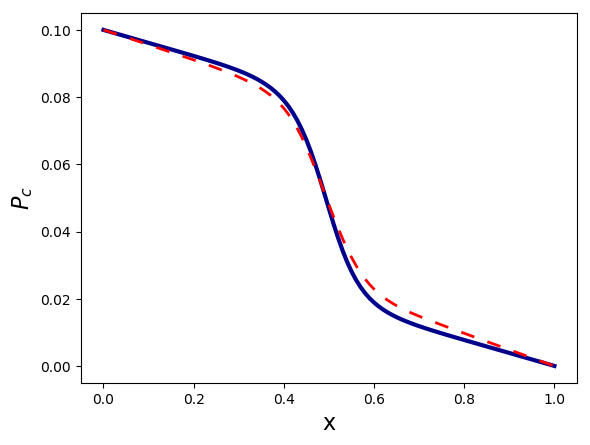}}}
	\caption{Comparison between physics-constrained DNN predictions and CFD solutions of idealized stenotic flows with three different stenosis geometries: (left) $A = 2\times10^{-3}$, (middle) $A = 4\times10^{-3}$, (right) $A = 7\times10^{-3}$. (a) streamwise velocity component $u$ (b) spanwise velocity component $v$ (c) centerline pressure profile $P_c$.}
	\label{fig:stenotic_geo}
\end{figure}
We compare the flow fields predicted by the trained DNN against the CFD benchmarks, and the results of three different samples of stenosis geometries, $A = 2\times{10^{-3}}$, $4\times{10^{-3}}$, and $7\times{10^{-3}}$, are shown in Fig.~\ref{fig:stenotic_geo}. From left to right, the degree of stenosis increases and thus the total flow rate is reduced due to the increased resistance (Fig.~\ref{fig:stenotic_geo}a and Fig.~\ref{fig:stenotic_geo}b). The pressure drop becomes more nonlinear as the stenotic vessel turns to be narrower (Fig.~\ref{fig:stenotic_geo}c). We can see that the DNN predictions can capture these flow features and agree with the CFD benchmarks well. Admittedly, the streamwise velocity of the flow with the narrowest stenotic vessel is slightly underestimated, and the centerline pressure profiles predicted by the DNN and CFD has a small discrepancy. This might be because the increased nonlinearity due to the steep geometric variation poses a challenge on the learning.
 
\begin{figure}[htb]
  \centering
    \subfloat[Viscosity uncertainty propagation]{\includegraphics[width=0.39\textwidth]{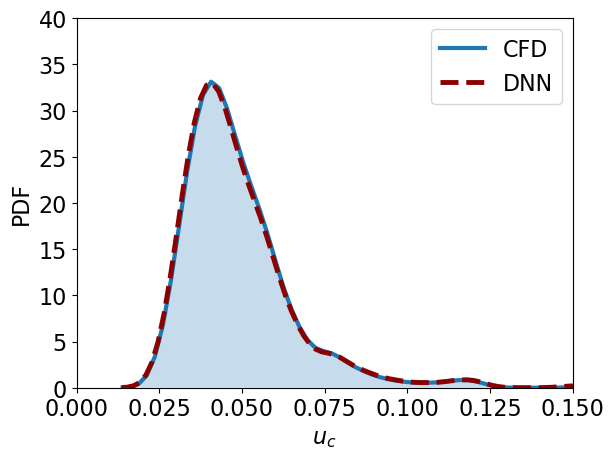}}
    \subfloat[Geometry uncertainty propagation] {\includegraphics[width=0.38\textwidth]{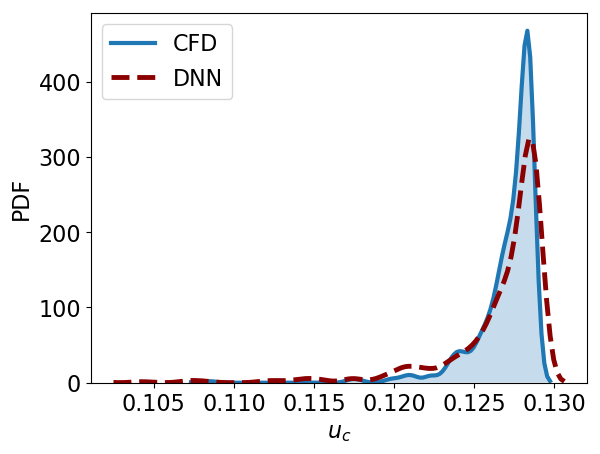}}
    \caption{Probability density of the center velocity $u_c$ propagated from (a) a normally distributed viscosity $\nu$ and (b) a normally distributed geometric parameter $A$ based on the trained DNN surrogate, compared against CFD-based MC solutions.}
    \label{fig:UQ_stenotic}
\end{figure}

After the training, the DNN surrogates are used to rapidly propagate uncertainties in viscosity and vessel geometry, and the effects on QoIs are investigated. Specifically, 500 MC samples are drawn from a normally distributed viscosity $\nu$ and geometric parameter $A$, which are propagated to the center velocities $u_c$ (at $x=0.5, y=0.0$) through the DNN surrogates. The probability distributions of $u_c$ due to uncertain viscosity and vessel geometry are shown in Figs.~\ref{fig:UQ_stenotic}a and ~\ref{fig:UQ_stenotic}b, respectively, where the propagated results through the CFD solver are also plotted for comparison.   
It shows that the propagated distributions of $u_c$ are non-Gaussian in both cases, which is due to the strong nonlinearity of the Navier--Stokes operator. As expected, the DNN-propagated uncertainties present a good agreement with the CFD-based benchmarks, especially in the case of viscosity uncertainty propagation (Figs.~\ref{fig:UQ_stenotic}a), where the two PDF curves are almost overlapped with each other. As for the geometry uncertainty propagation, although the overall feature of the PDF is captured, the peak density is slightly underpredicted by the DNN. The reason behind this could be that the DNN surrogate tends to underestimate the velocity magnitude in particular for a steep geometry variation.  

\subsubsection{Flow in Idealized Aneurysm}
We first learn the aneurysmal flows with varying viscosity, where the geometry of the aneurysm is fixed ($A = -5\times{10^{-3}}$).  
\begin{figure}[htb]
	\centering 
	\sidesubfloat[]{\fbox{\includegraphics[width=0.28\textwidth]{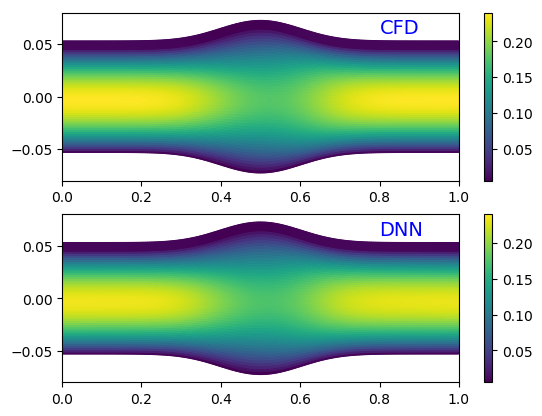}
	\includegraphics[width=0.28\textwidth]{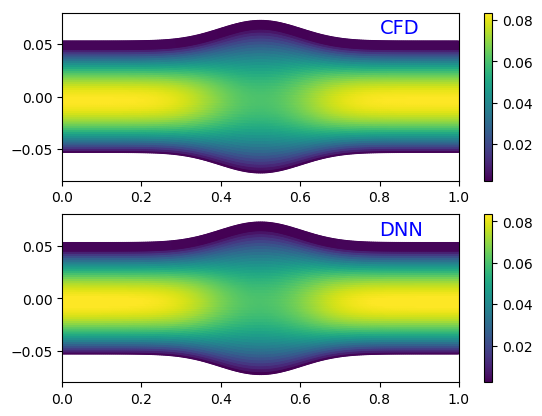}
	\includegraphics[width=0.28\textwidth]{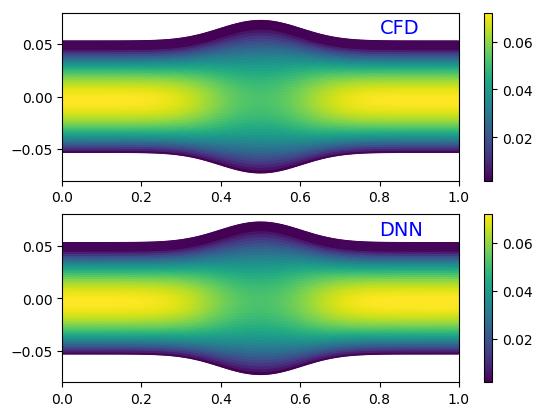}}}
	\par\medskip
	\vfill
	\sidesubfloat[]{\fbox{\includegraphics[width=0.28\textwidth]{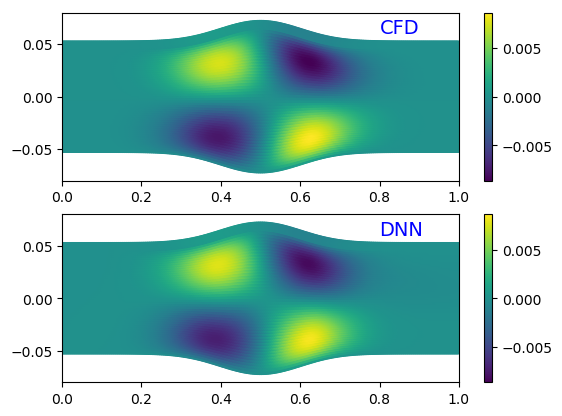}
	\includegraphics[width=0.28\textwidth]{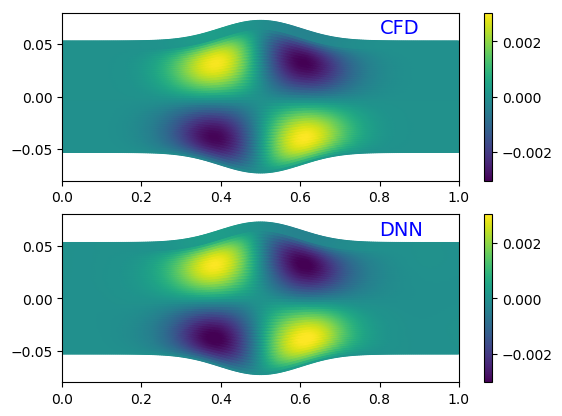}
	\includegraphics[width=0.28\textwidth]{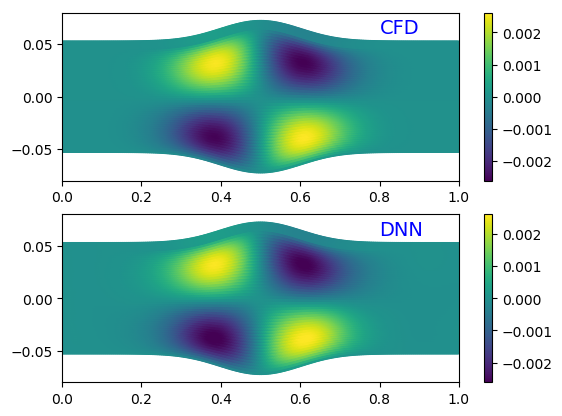}}}
	\par\medskip
	\vfill
	\sidesubfloat[]{\fbox{\includegraphics[width=0.28\textwidth]{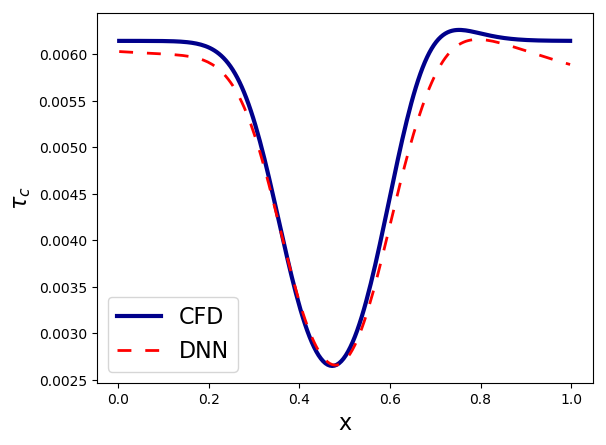}
	\includegraphics[width=0.28\textwidth]{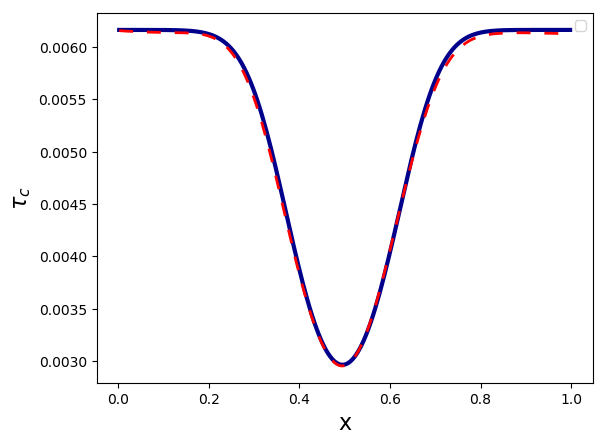}
	\includegraphics[width=0.28\textwidth]{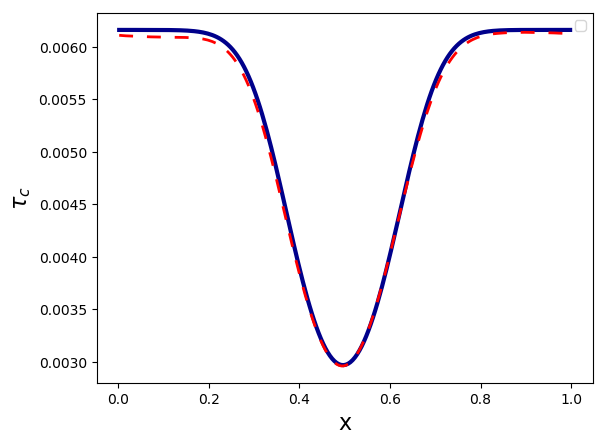}}}
	\caption{Comparison between physics-constrained DNN predictions and CFD solutions of idealized aneurysmal flows at three different viscosity samples: (left) $\nu = 6.4\times10^{-4}$, (middle) $\nu = 1.85\times10^{-3}$, (right) $\nu = 2.14\times10^{-3}$. (a) streamwise velocity component $u$ (b) spanwise velocity component $v$ (c) centerline wall shear profile $\tau_c$.}
	\label{fig:aneurysmal_vis}
\end{figure}
Training is conducted by sampling the viscosity points ranging from $5\times 10^{-4}$ to $1\times10^{-2}$. Figure~\ref{fig:aneurysmal_vis} shows the DNN-predicted flow fields of three viscosity samples, i.e, $\nu = 6.4\times10^{-4}, 1.85\times{10^{-3}}$, $2.14\times10^{-3}$, where the CFD benchmarks are plotted for comparison. In addition to the flow velocity and pressure, here we also investigate the wall shear stress (WSS) $\tau$, which has been demonstrated as a critical factor affecting the aneurysm initialization, progression, and rupture~\cite{chalouhi2013review}. It can be seen from Fig.~\ref{fig:aneurysmal_vis} that the DNN-predicted flow QoIs are in a good agreement with the CFD solutions. The decrease of the flow velocity and the minimum WSS of the vessel are accurately captured by the DNN surrogate.

\begin{figure}[htb]
	\centering 
	\sidesubfloat[]{\fbox{\includegraphics[width=0.28\textwidth]{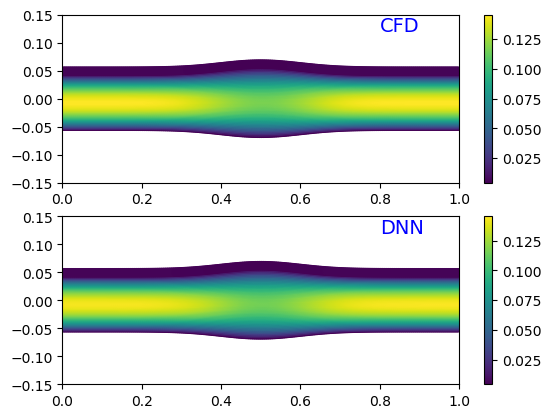}
	\includegraphics[width=0.28\textwidth]{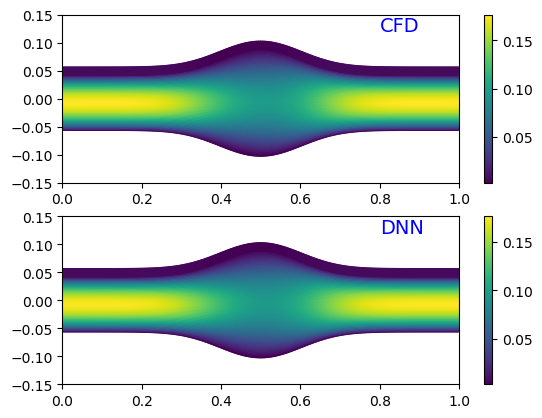}
	\includegraphics[width=0.28\textwidth]{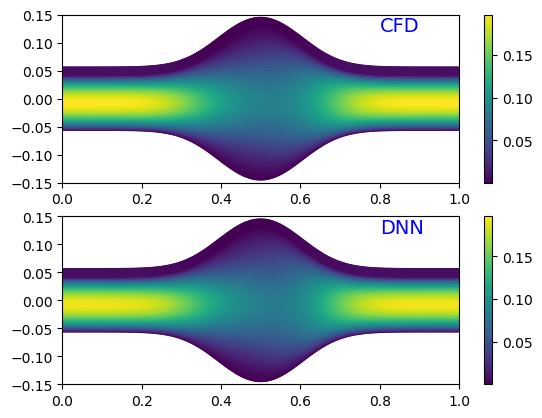}}}
	\par\medskip
	\vfill
	\sidesubfloat[]{\fbox{\includegraphics[width=0.28\textwidth]{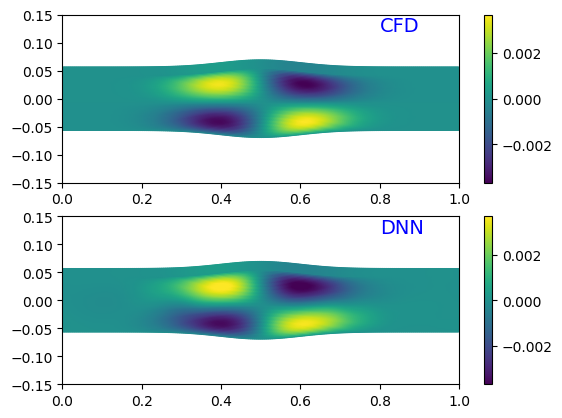}
	\includegraphics[width=0.28\textwidth]{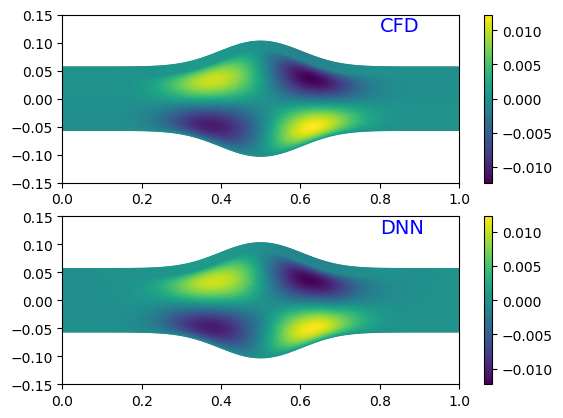}
	\includegraphics[width=0.28\textwidth]{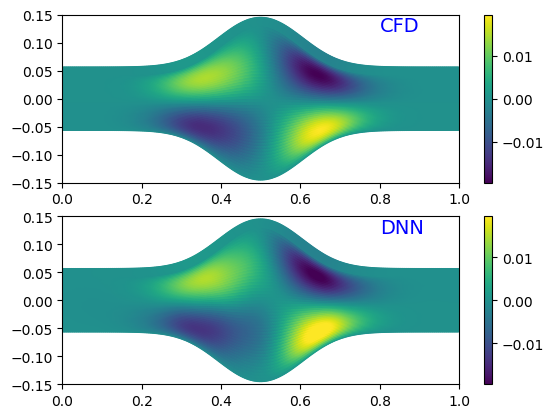}}}
	\par\medskip
	\vfill
	\sidesubfloat[]{\fbox{\includegraphics[width=0.28\textwidth]{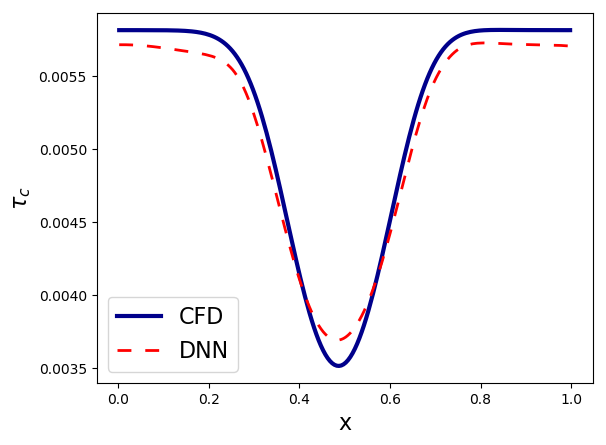}
	\includegraphics[width=0.28\textwidth]{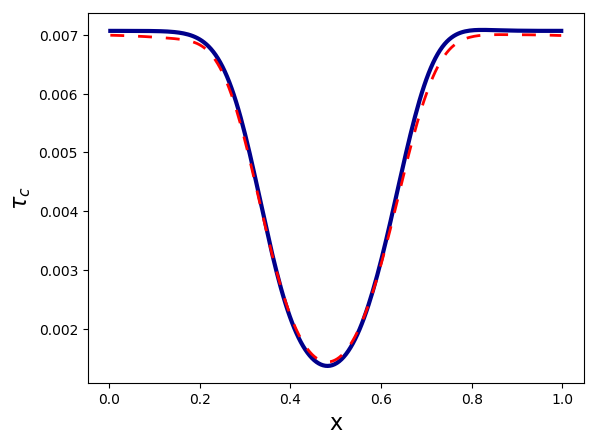}
	\includegraphics[width=0.28\textwidth]{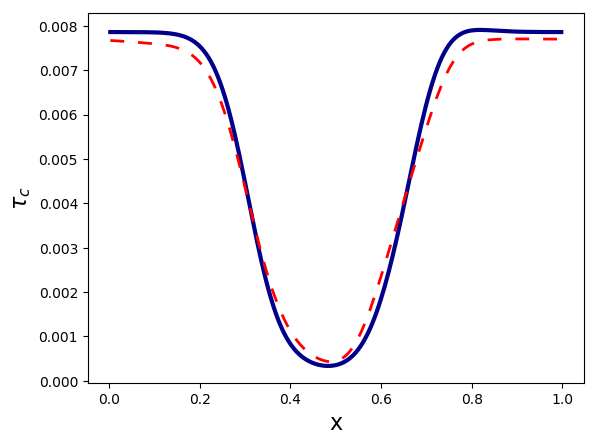}}}
	\caption{Comparison between physics-constrained DNN predictions and CFD solutions of idealized aneurysmal flows of three different aneurysm geometries: (left) $A = -3\times10^{-3}$, (middle) $A = -1.2\times10^{-2}$, (right) $A = -2.2\times10^{-2}$. (a) streamwise velocity component $u$ (b) spanwise velocity component $v$ (c) centerline wall shear profile $\tau_c$. }
	\label{fig:aneurysmal_geo}
\end{figure}
Figure~\ref{fig:aneurysmal_geo} shows the performance for learning the geometry-varying solutions of the aneurysmal flow, where the training is conducted by sampling the geometric parameter $A$, ranging from $-2\times10^{-2}$ to $0$. The predicted flow fields of three different geometry samples are presented, where the size of the aneurysm increases from left to right. The flow decelerates through the expanded region of the vessel and the velocity at the center of the aneurysm is significantly reduced, in particular when the aneurysm becomes larger. It is observed from the contour comparisons (Figs.~\ref{fig:aneurysmal_geo}a and~\ref{fig:aneurysmal_geo}b), the DNN predictions agree with the CFD solutions pretty well. As for the WSS profile, its shape and magnitude vary as the geometry changes, which can be accurately captured by the DNN surrogate (Figs.~\ref{fig:aneurysmal_geo}c). 

The uncertainty propagation using the trained DNN surrogates is then conducted. 
\begin{figure}[htb]
\centering
\subfloat[Viscosity uncertainty propagation]{\includegraphics[width=0.39\textwidth]{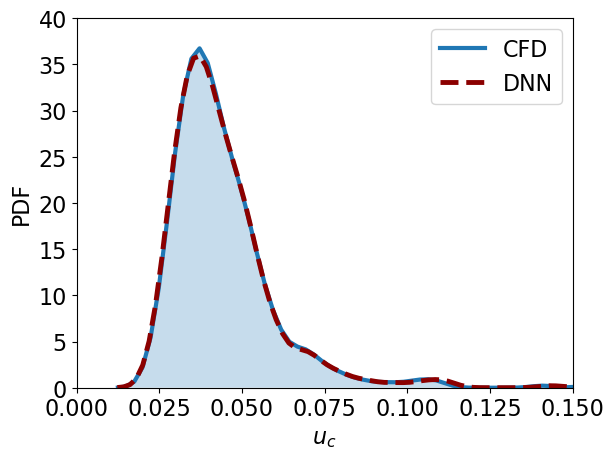}}
\subfloat[Geometry uncertainty propagation]{\includegraphics[width=0.38\textwidth]{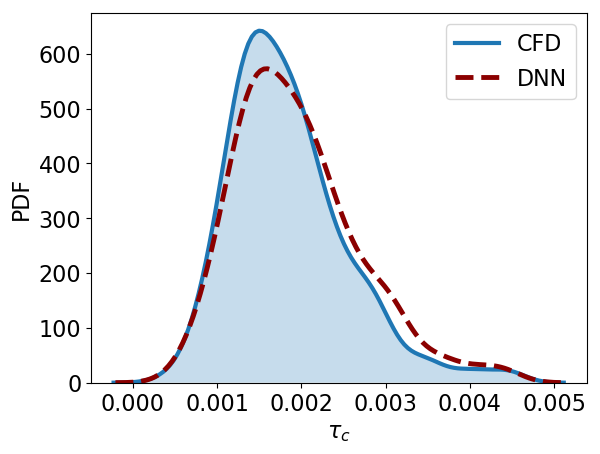}}
\caption{Probability density of the (a) center velocity $u_c$ propagated from  a normally distributed viscosity $\nu$ and (b) the minimum wall shear $\tau_c$ propagated from a normally distributed geometric parameter $A$ based on the trained DNN surrogate, compared against CFD-based MC solutions.}
\label{fig:UQ_aneurysmal}
\end{figure}
Uncertainties in viscosity $\nu$ and geometric parameter $A$ with Gaussian distributions are considered, and the QoIs are center velocity $u_c$ and the minimum WSS $\tau_c$, which are important for aneurysmal flows. As shown in Fig.~\ref{fig:aneurysmal_vis}c, the WSS remains invariant due to viscosity perturbation with a fixed vessel geometry, hence the distribution of $u_c$ is studied for viscosity uncertainty propagation (shown in Fig.~\ref{fig:UQ_aneurysmal}a). The PDF obtained by the DNN surrogate almost coincides with the CFD-based benchmark. As for the geometry uncertainty propagation, the center velocity $u_c$ is not an interesting quantity since it almost remains the same as geometry changes. Instead, we investigate the propagated uncertainty in the minimum WSS $\tau_c$, which is sensitive to geometry variation. It is observed from Fig.~\ref{fig:UQ_aneurysmal}b that the probabilistic distribution of $\tau_c$ propagated by the DNN is in a favorable agreement with the CFD-based benchmark, though the peak of the density is slightly underestimated.         

\subsubsection{Summary of Training and Prediction Performance}
The training and testing performance on the vascular flow cases are summarized in Table~\ref{tab:evasummary}, where the training loss is defined as the sum of $L_2$ norms of momentum and continuity equation residuals (see eq.~\ref{eq:op3}) and test error is defined as the normalized $L_2$ difference between DNN and CFD results, shown as,
\begin{subequations}
	\label{eq:udiff}
	\begin{alignat}{2}
    e_{\mathbf{u}} = \frac{||\mathbf{u}_{DNN}- \mathbf{u}_{CFD}||_{2}}{\Delta p}\\
    e_{p} = \frac{||{p}_{DNN}- {p}_{CFD}||_{2}}{(\Delta p)^2}
    \end{alignat}
\end{subequations}
where pressure drop $\Delta p = 0.1$ is used to normalize the errors. 

\begin{table}[htp]
\begin{center}
\begin{tabular}{ |c|cccc| } 
 \hline
 \multirow{2}{*}{}&\multicolumn{2}{c}{Stenosis}&\multicolumn{2}{c|}{Aneurysm} \\
 &Viscosity & Geometry &Viscosity&Geometry \\
\hline
Training Loss &$8\times10^{-5}$&$1\times10^{-3}$&$5.5\times10^{-5}$&$2.9\times10^{-5}$\\ 
Test error $e_{u}$ &$8.18\times10^{-5}$&$9.61\times10^{-4}$&$9.2\times10^{-5}$&$1.38\times10^{-4}$\\
Test error $e_{v}$ &$7.14\times10^{-8}$&$1.76\times10^{-6}$&$1.33\times10^{-7}$&$1.15\times10^{-6}$\\
Test error $e_{p}$ &$2.33\times10^{-5}$&$2.23\times10^{-3}$&$5.81\times10^{-6}$&$2.09\times10^{-5}$\\
\hline
\end{tabular}
\end{center}
\caption{Summary of learning/prediction performance of vascular flows with $20$ parameter collocation points in parameter spaces for training}
\label{tab:evasummary}
\end{table}

It can be seen from Table~\ref{tab:evasummary} that both the training loss and test errors are reasonably small for all cases after training, and the test error in $u$ is more dominant than the others. Among the four scenarios, the geometric variation of the stenosis is the most challenging to learn since the training loss (i.e., equation residual) remains relatively large and the test errors are one-order bigger than the other cases. This is consistent with the notable discrepancy in the DNN-predicted PDF observed in Fig.~\ref{fig:UQ_stenotic}b.       

\begin{figure}[htb]
\centering
\includegraphics[width=0.5\textwidth]{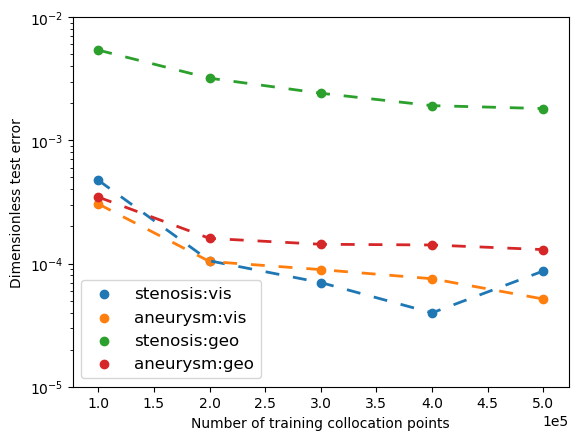}
\caption{Total test errors v.s. different number of training collocation points. The total collocation points is the multiplication of parameter point size ($N_p$) by geometric collocation point size ($N_g = 10^4$).}
\label{fig:avg_lossl}
\end{figure}

All the cases presented above are trained on $N_p = 20$ parameter collocation points. Namely, the equation residuals are minimized on $20$ collocation points uniformly sampled from the parameter space. It is necessary to check if the size $N_p$ of parameter collocation points is sufficiently large for the training. Therefore, we conduct a parameter study using different amounts of parameter collocation points for training. The total test errors (sum of test errors in $u, v$, and $p$) against different numbers of training collocation points for all four cases are presented in Fig.~\ref{fig:avg_lossl}.  
As expected, the errors generally decrease as the number of training collocation points increases. However, the error decreasing rate is quite mild. The test errors remain approximately unchanged when $N_p > 20$ for all cases while the training cost will increase as more collocation points are used, which justifies the sufficiency of the total collocation points used in this work. Detailed results of training cost and testing errors for all cases are summarized by Tables~\ref{tab:apptab1} and~\ref{tab:apptab2} in Appendix A.

\subsubsection{Computational Cost of Uncertainty Propagation}
Finally, we briefly discuss the computational cost of the uncertainty propagation tasks presented above, which is determined by the forward model-evaluation time and number of MC samples. For both stenotic and aneurysmal flows, a structured mesh with $10^4$ quadrilateral grids is used for the CFD model. The computational cost of each forward solution of the fully-converged CFD simulation is around 40 CPU seconds, and propagating 500 MC samples in test cases takes about 20,000 CPU seconds on a single CPU core. However, the online evaluation of the trained DNN surrogate is very fast, and the cost of each forward DNN evaluation is less than $2\times10^{-2}$ CPU seconds. Therefore, the propagation of 500 MC samples by the DNN surrogate only takes about 10 CPU seconds, which gains 2000-times speedup over the CFD-based uncertainty propagation. This advantage can be considerable when a larger number of forward evaluations are needed or more complicated fluid systems are considered. Moreover, the FV/FE-based CFD simulations require mesh generation, which is often a manually-cumbersome and labor-intensive process, in particular for the flows with complex geometries and moving boundaries, e.g, patient-specific cardiovascular simulations. Therefore, the mesh-free feature of the proposed method shows great promises for the many-query analysis in these systems. 

\section{Discussion}
\label{sec:discussion}
\subsection{Pitfall on Using Soft Boundary Enforcement}
\begin{figure}[htp]
	\centering 
	\subfloat[CFD benchmark]{\includegraphics[width=0.33\textwidth]{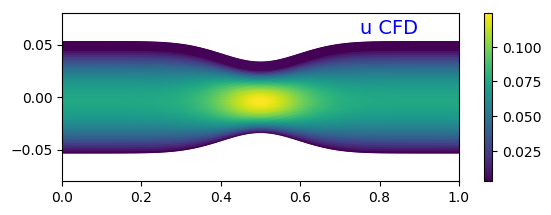}}
	\subfloat[DNN hard]{\includegraphics[width=0.33\textwidth]{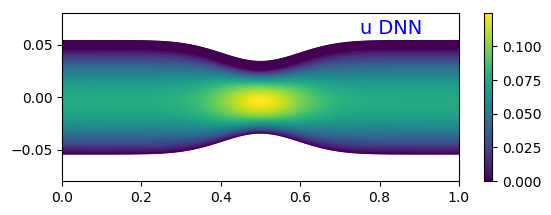}}
    \subfloat[DNN soft $\lambda = 1$]{\includegraphics[width=0.33\textwidth]{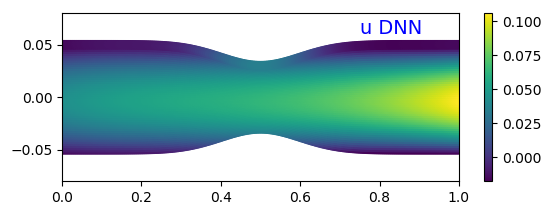}}
	\vfill
	\subfloat[DNN soft $\lambda = 10$]{\includegraphics[width=0.33\textwidth]{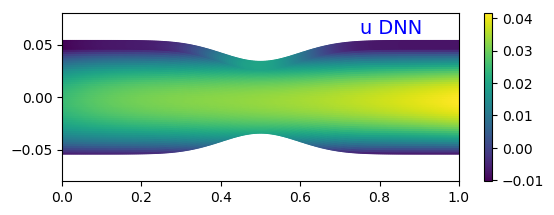}}
	\subfloat[DNN soft $\lambda = 100$]{\includegraphics[width=0.33\textwidth]{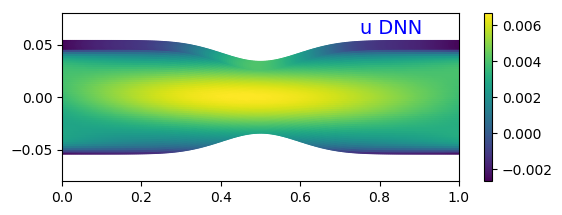}}
	\subfloat[DNN soft $\lambda = 1000$]{\includegraphics[width=0.33\textwidth]{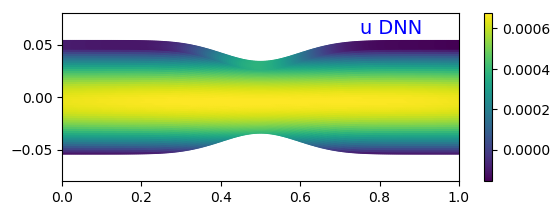}}
	\caption{Physics-constrained learning results for a stenotic flow ($A = 5\times 10^{-3}, \nu = 1\times10^{-3}$) with (b) hard BC constraint, compared to those using soft BC constraints with (c) $\lambda = 1$, (d) $\lambda = 10$, (e) $\lambda = 100$, (f) $\lambda = 1000$.}
\label{fig:draft}
\end{figure}

The initial and boundary conditions (IC/BC) can be imposed in the physics-constrained learning either as a soft or hard constraint. When no labeled data is used in training, a properly enforced IC/BC is crucial to ensure the uniqueness of the learned PDE solutions. Although we have demonstrated the effectiveness of the hard enforcement approach in Sec.~\ref{sec:result}, it is still interesting to investigate the performance of soft enforcement method~\cite{raissi2019physics} in the purely PDE-driven training. Hence, all the test flows are studied again using physics-constrained learning, where BCs are imposed in a soft manner (as Eq.~\ref{eq:op4}). Namely, the BCs are formulated as a boundary loss component $\mathcal{L}_{B}$, which is incorporated into the physics-based loss function $\mathcal{L}_{phy}$ as,
\begin{equation}
\label{eq:op4}
    \mathcal{L}_{phy}^c = \mathcal{L}_{phy} + \lambda\mathcal{L}_{B},
\end{equation}
where $\lambda$ is the penalty coefficient. For the circular pipe flow case, both the hard and soft BC constraints can lead to excellent learning and prediction performance (results not shown). However, when the radius varies along the tube as in the stenosis and aneurysm cases, the DNN with the soft BC enforcement does not perform well as the no-slip BC of the vessel wall is poorly imposed especially near the bottleneck in Fig.~\ref{fig:draft}. Consequently, the solution to the flow field becomes inaccurate. For example, Fig.~\ref{fig:draft} shows the results for learning a stenotic flow ($A = 5\times 10^{-3}$) using the soft constraint with different $\lambda$ values, where the result with hard boundary constraints and CFD benchmark are plotted as well for comparison.  
In contrast to the result with hard constraints (Fig.~\ref{fig:draft}b), both the flow patterns and magnitudes predicted with the soft boundary enforcement with different $\lambda$ (Fig.~\ref{fig:draft}c--f) are completely wrong compared to the CFD benchmark (Fig.~\ref{fig:draft}a). Moreover, the nonlinear behavior of the centerline pressure profile cannot be accurately captured (results not shown). 
\begin{table}[htp]
\begin{center}
\begin{tabular}{ |c|c|c|c|c| } 
 \hline
 Loss & $\lambda = 1$ & $\lambda = 10$& $\lambda = 100$ & $\lambda = 1000$\\ 
 \hline
 Boundary Condition & $1\times10^{-3}$ &$2\times10^{-4}$&$7\times10^{-6}$ & $1\times10^{-7}$ \\
 \hline
 $x$-Momentum Equation & $1\times10^{-4}$&$3\times10^{-3}$&$8\times10^{-3}$ & $1\times10^{-2}$ \\ 
 \hline
\end{tabular}
\end{center}
\caption{Converged loss with different penalty coefficient $\lambda$}
\label{tab:softloss}
\end{table}
The poor performance reflects the major drawbacks of the soft constraints in the physics-driven training as mentioned above. First, unique PDE solutions are determined by the IC/BC, which cannot be guaranteed by simply penalizing the boundary loss. Moreover, it is difficult to assign the relative weight ($\lambda$) for different components in the loss function, and there could be ``competing" effects between the equation loss and boundary loss, which makes the optimization difficult to converge. As shown in Table~\ref{tab:softloss}, by assigning a large weight for the boundary loss ($\lambda = 1000$), the boundary condition can be well prescribed but the PDE residual remains a large value and cannot be further reduced. On the other hand, when the weight for the boundary is small ($\lambda = 1$), the loss of x-momentum equation can be reduced to $O(10^{-4})$ but the BC fails to be imposed. It is important to note that none of these four $\lambda$ leads to a physical stenotic flow pattern as shown in Fig.~\ref{fig:draft}.  

\subsection{Role of Activation Function in Physics-Constrained Learning}
The performance of DNN training is affected by the activation function to a large degree. The widely used activation functions includes ReLU, Sigmoid, Tanh, etc.~\cite{maas2013rectifier}. However, these activation functions are not guaranteed to be optimized in terms of the convergence rate and accuracy. Recent studies~\cite{ramachandran2017searching,jagtap2019adaptive} proposed to train an adaptive activation function as well as the neural network weights to achieve better convergence property. Notably, Ramachandran et al.~\cite{ramachandran2017searching} introduced an adaptive activation function called Swish, which is defined as $x\cdot \textrm{Sigmoid}(\beta x)$ and $\beta$ is a trainable parameter. Jagtap and Karniadakis~\cite{jagtap2019adaptive} presented a new adaptive function defined as $\textrm{Tanh}(nax)$, where $a$ is an adaptive parameter to be learned and $n$ is a scale factor that potentially speeds up the convergence. 

\begin{figure}[htp]
	\centering 
	\subfloat[Hard BC enforcement]{\includegraphics[width=0.45\textwidth]{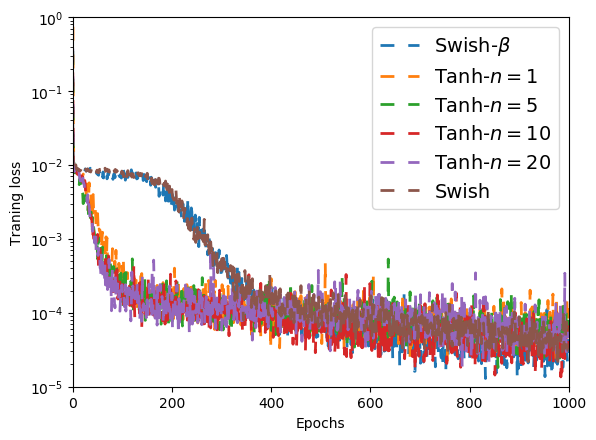}}
	\subfloat[Soft BC enforcement]{\includegraphics[width=0.45\textwidth]{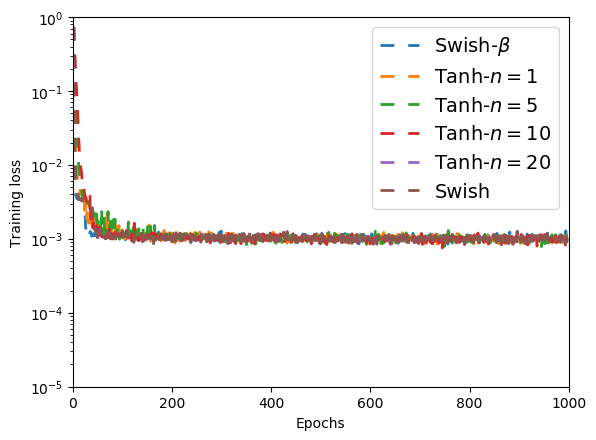}}
	\caption{Learning curves of different activation functions on a stenotic flow ($A = 5\times 10^{-3}, \nu = 1\times10^{-3}$) with (a) hard BC constraint, (b) soft BC constraint with the penalty coefficient $\lambda = 1$}
\label{fig:convergence}
\end{figure}

In current work, the training process uses a Swish activation function with fixed $\beta = 1$. It's not clear how the adaptivity of activation function can affect the convergence rate and accuracy. Furthermore, as we discussed and highlighted the necessity of the hard BC enforcement in the physics-constrained data-free learning, it is also interesting to compare the relative importance of boundary condition enforcement and adaptive activation function on model performance. Therefore, we test the effects of different activation functions with/without hard boundary constraints in the stenosis case ($A = 5\times10^{-3}$, $\nu = 1\times10^{-3}$), and the resulting learning curves are shown in Fig.~\ref{fig:convergence}, where panels~(a) and~(b) show the convergence histories of different activation functions with and without a hard BC enforcement, respectively. 
The legend ``Swish-$\beta$" refers to the Swish function with adaptive $\beta$, while the ``Swish" means the Swish function with a fixed $\beta = 1$. The results of adaptive Tanh activation function with different $n$~\cite{jagtap2019adaptive} are plotted as well. It can be seen in Fig.~\ref{fig:convergence}a that the convergence rate of using adaptive Tanh functions is faster than that of the Swish activation functions within the first 400 epochs, but the training loss of all cases finally converge to the same order. Moreover, the convergence curves almost overlap with each other for the same type of activation function with different hyperparameters. When the BC is imposed in a soft manner (Fig.~\ref{fig:convergence}b), the solution can be quickly trapped in a bad local minimum as we discussed above and the final training loss is over one-order larger than that of the cases with hard BC constraints. Using different adaptive activation functions does not help to further decrease the loss and all the convergence curves are overlapped. Moreover, the convergence history of each trainable hyperparameter (i.e., $\beta$ in Swish function and $a$ in Tahn function) is monitored (see Fig.~\ref{fig:adpAc} in Appendix B), and we found that these hyperparameters are more likely to converge to the optimized values when the BCs are enforced in a hard manner. To sum up, the way of imposing BCs is found to be more critical than the adaptivity of activation function in our cases in terms of accuracy in the physics-constrained data-free learning.

\subsection{Physics-Constrained Data-Free Learning vs. Traditional Data-Driven Learning}
We have demonstrated that the solutions of parametric Navier-Stokes equations can be effectively learned by solely minimizing the PDE residuals without using any simulation data. To better evaluate the advantages and limitations of the physics-constrained learning, we also conducted a series of comparison studies between the physics-constrained data-free DNN and traditional data-driven DNN in terms of learning efficiency and prediction accuracy. Namely, all the vascular flow cases discussed above are learned again in a purely data-driven way, where the DNN architecture and hyperparameters remain the same and only the physics-based loss function (i.e., Eq.~\ref{eq:op3}) is replaced with the data-based one (i.e., Eq.~\ref{eq:op1}). A number of CFD simulations with different input parameters are conducted and the simulated velocity and pressure fields are collected as the training data. The detailed comparisons for stenotic and aneurysmal flows cases with varying viscosity and geometry are summarized by Tables~\ref{tab:apptab1} and~\ref{tab:apptab2} in Appendix A. In general, the prediction results from purely data-driven learning are slightly more accurate than those of physics-constrained learning, and the accuracy of both models improves as the parameter collocation points increase. However, data-driven learning requires additional offline CFD simulations and this computational overhead can quickly grow as more training points are sampled from the parameter space. In this paper, since the CFD cases considered here are not costly to simulate (e.g., each CFD simulation takes about 40 CPU seconds), the computational overhead due to the offline data generation process is not significant. Nonetheless, the advantage of the data-free feature in physics-constrained learning will become more notable when large-scale 3D flow problems are considered, where a single simulation run could be very expensive. It notes that when training data are ready for use, the cost of data-driven training process is approximately similar to that of the physics-constrained training (in the same order), though the latter is slightly slower due to the additional AD calculations for derivatives. To reduce the training cost in data-driven learning, one way is to reduce the spatial dimensionality by projecting the training data (i.e., velocity/pressure fields) onto the POD basis, and learning is performed on POD coefficients instead of spatial collocation points. We also conducted POD-based data-driven learning and found that the learning performance with the current shallow network structure is unsatisfactory (results are not shown here for conciseness). 

Admittedly, the current form of physics-constrained DNN has its limitations, for example, the offline training process is still costly, the convergence cannot be guaranteed due to the non-convexity of DNN optimization, and scalability for high-dimensional complex problems is still challenging. The proposed PDE-constrained DNN is not expected to replace the classical CFD (numerical) solvers, which have been developed for decades. However, the development of PDE-constrained DNN for surrogate modeling shows strong promise. Particularly, the proposed method is mesh-free and thus does not require arduous mesh generation labor and intensive domain expertise in numerical modeling, which is suitable for, e.g., rapidly testing ideas in the design phase. We expect that the effectiveness of surrogate model based on physics-constrained DNN will be significantly promoted along with the rapid advances in improvement of DNN training efficiency, e.g., a recent study has suggested that a novel photonic chip has a potential to be used to train deep neural networks 10-million of times more efficiently than current CPUs/GPUs do~\cite{hamerly2019large}.

\section{Conclusion}
\label{sec:conclusion}
Surrogate modeling of fluid flows governed by the Navier--Stokes equations is significant for uncertainty quantification, optimization design, and inverse analysis in many engineering systems. As a universal function approximator, DNN is becoming a popular approach for surrogate modeling. However, training of a DNN often requires large number of labeled data, which are usually not available for efficiently developing surrogates since each data point requires an expensive CFD simulation. This paper presented a novel DNN surrogate for fluid simulations without using any labeled data (i.e., CFD simulation data). Specifically, a structured DNN architecture is devised to approximate the solutions of the parametric Navier--Stokes equations, where the initial/boundary conditions are satisfied automatically. Instead of using any simulation data, the DNN is trained by solely minimizing the violation of the mass and momentum conservation laws for fluid flows. Compared to the previous works of physics-constrained learning, this paper focuses on modeling of fluid systems governed by parametric Navier--Stokes equations. The proposed methods were tested on three flow cases relevant to cardiovascular applications, i.e., circular pipe flow, stenotic flow, and aneurysmal flow. The DNNs with equation-based loss were trained to learn the flow fields with parameter variations in, e.g., viscosity and domain geometry. Uncertainties in these parameters are propagated through the trained DNN surrogate and the results are validated against the CFD benchmarks. The comparisons indicate the excellent agreement between the physics-constrained DNN surrogate models and CFD simulations. Without using any labeled data in training, the DNN is able to accurately parameterize the velocity/pressure solutions with varying viscosity and geometries, which can be used to efficiently propagate uncertainties with enormous MC samples. Moreover, the performances of using hard and soft IC/BC enforcement approaches are compared and the issues of soft constraints in physics-constrained are discussed. We also investigated the influence of state-of-art adaptive activation functions and compared the present labeled-data-free learning approach with traditional data-driven learning approach in terms of accuracy and efficiency. In summary, the results have demonstrated the merit of the proposed method and suggest a great promise in developing DNN for surrogate fluid models without the need for CFD simulation data.   
         
\section*{Compliance with Ethical Standards}
The authors declare that they have no conflict of interest.

\section*{Acknowledgement}
LS gratefully acknowledge partial funding of graduate fellowship from China Scholarship Council (CSC) in this effort. JXW would acknowledge support from the Defense Advanced Research Projects Agency (DARPA) under the Physics of Artificial Intelligence (PAI) program (contract HR00111890034). The authors also would like to thank Dr. Nicholas Zabaras and Mr. Yinhao Zhu for their helpful discussions during this work.

\section*{Appendix A}
\label{sec:ap1}
The physics-constrained, data-free model and purely data-driven model are trained on a single NVIDIA 1080TI GPU card.
\begin{table}[htp]
\footnotesize{
\begin{center}
\begin{tabular}{ |c|ccccc| } 
\hline
\multicolumn{6}{|c|} {\textbf{Stenosis with varying viscosity (physics-constrained data-free)}}\\
\hline
Training viscosity points &10&20&30&40&50\\ 
\hline
Training cost on GPU& 7598s&12977s&25988s&28543s&31996s\\
\hline
Training minibatch loss&$3.11\times 10^{-4}$&$8\times 10^{-5}$&$7.64\times 10^{-6}$&$4.6\times 10^{-5}$&$5.8\times 10^{-5}$\\
\hline
Test error $e_u$&$4.13\times 10^{-4}$&$8.18\times10^{-5}$&$6.02\times10^{-5}$&$3.29\times10^{-5}$&$7.27\times 10^{-5}$\\
Test error $e_v$ &$3.32\times10^{-7}$&$7.14\times10^{-8}$&$6.43\times10^{-8}$&$3.86\times10^{-8}$&$1.41\times 10^{-7}$\\
Test error $e_p$ &$6.08\times10^{-5}$&$2.33\times10^{-5}$&$9.82\times10^{-6}$&$6.87\times10^{-6}$&$1.41\times10^{-5}$\\
\hline
\multicolumn{6}{|c|} {\textbf{Stenosis with varying viscosity (purely data-driven)}}\\
\hline
Training viscosity points &10&20&30&40&50\\ 
\hline
Training cost on GPU&6900s&11967s&19982s&22497s&30893s\\
Simulation cost on CPU&400s&800s&1200s&1600s&2000s\\
\hline
Training minibatch loss&$5\times10^{-7}$&$5\times10^{-7}$&$1\times10^{-7}$&$1\times10^{-7}$&$1\times10^{-7}$\\
\hline
Test error $e_u$&$1.49\times10^{-5}$&$6.83\times10^{-6}$&$2.43\times10^{-6}$&$1.33\times10^{-6}$&$2.64\times10^{-6}$\\
Test error $e_v$& $7.84\times10^{-8}$&$2.37\times10^{-7}$&$8.84\times10^{-8}$&$3.09\times10^{-8}$&$3.01\times10^{-8}$\\
Test error $e_p$& $6.42\times10^{-7}$&$7.13\times10^{-7}$&$1.56\times10^{-6}$&$3.72\times10^{-7}$&$1.65\times10^{-6}$\\
\hline
\multicolumn{6}{|c|} {\textbf{Stenosis with varying geometry (physics-constrained data-free)}}\\
\hline
Training viscosity points &10&20&30&40&50\\ 
\hline
Training cost on GPU& 6458s&11688s&18239s&24674s&42814s\\
\hline
Training minibatch loss&$5.0\times 10^{-4}$&$1.0\times 10^{-3}$&$3.0\times 10^{-4}$&$7.0\times 10^{-4}$&$1.0\times 10^{-4}$\\
\hline
Test error $e_u$ &$1.54\times 10^{-3}$&$9.61\times10^{-4}$&$5.24\times10^{-4}$&$3.46\times10^{-4}$&$4.81\times10^{-4}$\\
Test error $e_v$ &$2.93\times10^{-6}$&$1.76\times10^{-6}$&$1.26\times10^{-6}$&$7.43\times10^{-7}$&$9.53\times 10^{-7}$\\
Test error $e_p$ &$3.85\times10^{-3}$&$2.23\times10^{-3}$&$1.89\times10^{-3}$&$1.57\times10^{-3}$&$1.33\times10^{-3}$\\
\hline
\multicolumn{6}{|c|} {\textbf{Stenosis with varying geometry (purely data-driven)}}\\
\hline
Training viscosity points &10&20&30&40&50\\ 
\hline
Training cost on GPU&5498s&11238s&16285s&21848s&26396s\\
Simulation cost on CPU&400s&800s&1200s&1600s&2000s\\
\hline
Training minibatch loss&$3\times10^{-7}$&$1\times10^{-7}$&$2\times10^{-7}$&$1\times10^{-7}$&$1\times10^{-7}$\\
\hline
Test error $e_u$&$3.21\times10^{-5}$&$2.18\times10^{-5}$&$2.4\times10^{-5}$&$2.74\times10^{-5}$&$3.12\times10^{-5}$\\
Test error $e_v$ & $2.11\times10^{-7}$&$1.37\times10^{-7}$&$6.92\times10^{-8}$&$9.33\times10^{-8}$&$5.21\times10^{-8}$\\
Test error $e_p$ & $2.72\times10^{-6}$&$2.88\times10^{-7}$&$1.22\times10^{-6}$&$3.41\times10^{-7}$&$6.40\times10^{-7}$\\
\hline
\end{tabular}
\end{center}
\caption{Training and testing performance for stenosis case with varying viscosity/geometry}
\label{tab:apptab1}
}
\end{table}

\begin{table}[htp]
\footnotesize{
\begin{center}
\begin{tabular}{ |c|ccccc| } 
\hline
\multicolumn{6}{|c|} {\textbf{Aneurysm with varying viscosity (physics-constrained data-free)}}\\
\hline
Training viscosity points &10&20&30&40&50\\ 
\hline
Training cost on GPU& 5731s&14490s&25350s&24810s&31595s\\
\hline
Training minibatch loss&$9.71\times 10^{-5}$&$5.50\times 10^{-5}$&$1.07\times 10^{-5}$&$2.33\times 10^{-5}$&$2.22\times 10^{-5}$\\
\hline
Test error $e_u$&$2.72\times 10^{-4}$&$9.2\times10^{-5}$&$8.33\times10^{-5}$&$6.4\times10^{-5}$&$4.67\times 10^{-5}$\\
Test error $e_v$ &$2.22\times10^{-7}$&$1.33\times10^{-7}$&$1.31\times10^{-7}$&$5.96\times10^{-8}$&$7.46\times 10^{-8}$\\
Test error $e_p$ &$6.14\times10^{-7}$&$1.18\times10^{-5}$&$5.81\times10^{-6}$&$1.12\times10^{-5}$&$4.75\times10^{-6}$\\
\hline
\multicolumn{6}{|c|} {\textbf{Aneurysm with varying viscosity (purely data-driven)}}\\
\hline
Training viscosity points &10&20&30&40&50\\ 
\hline
Training cost on GPU&6851s&11861s&24580s&24353s&31116s\\
Simulation cost on CPU&400s&800s&1200s&1600s&2000s\\
\hline
Training minibatch loss&$5\times10^{-7}$&$5\times10^{-7}$&$1\times10^{-7}$&$1\times10^{-7}$&$1\times10^{-7}$\\
\hline
Test error $e_u$&$1.76\times10^{-4}$&$5.11\times10^{-6}$&$8.51\times10^{-6}$&$5.30\times10^{-7}$&$3.20\times10^{-6}$\\
Test error $e_v$ & $1.07\times10^{-7}$&$6.31\times10^{-8}$&$1.56\times10^{-7}$&$6.95\times10^{-8}$&$7.42\times10^{-8}$\\
Test error $e_p$ & $6.14\times10^{-7}$&$2.38\times10^{-7}$&$1.81\times10^{-6}$&$1.06\times10^{-6}$&$3.28\times10^{-7}$\\
\hline
\multicolumn{6}{|c|} {\textbf{Aneurysm with varying geometry (physics-constrained data-free)}}\\
\hline
Training viscosity points &10&20&30&40&50\\ 
\hline
Training cost on GPU& 7755s&12904s&24645s&26724s&29515s\\
\hline
Training minibatch loss&$8.8\times 10^{-5}$&$2.9\times 10^{-5}$&$1.8\times 10^{-5}$&$7.8\times 10^{-6}$&$1.9\times 10^{-5}$\\
\hline
Test error $e_u$ &$3.07\times 10^{-4}$&$1.38\times10^{-4}$&$1.29\times10^{-4}$&$1.31\times10^{-4}$&$1.19\times10^{-4}$\\
Test error $e_v$ &$1.67\times10^{-6}$&$1.15\times10^{-6}$&$1.01\times10^{-6}$&$9.95\times10^{-7}$&$9.06\times 10^{-7}$\\
Test error $e_p$ &$3.77\times10^{-5}$&$2.09\times10^{-5}$&$1.28\times10^{-5}$&$8.99\times10^{-6}$&$1.02\times10^{-5}$\\
\hline
\multicolumn{6}{|c|} {\textbf{Aneurysm with varying geometry (purely data-driven)}}\\
\hline
Training viscosity points &10&20&30&40&50\\ 
\hline
Training cost on GPU&6322s&11607s&18582s&23309s&32124s\\
Simulation cost on CPU&400s&800s&1200s&1600s&2000s\\
\hline
Training minibatch loss&$3\times10^{-7}$&$2\times10^{-7}$&$4\times10^{-7}$&$1\times10^{-7}$&$8\times10^{-8}$\\
\hline
Test error $e_u$&$7.9\times10^{-5}$&$8.63\times10^{-5}$&$7.74\times10^{-5}$&$8.69\times10^{-5}$&$6.73\times10^{-5}$\\
Test error $e_v$ & $8.77\times10^{-7}$&$5.34\times10^{-7}$&$5.03\times10^{-7}$&$6.24\times10^{-7}$&$6.97\times10^{-7}$\\
Test error $e_p$ & $1.06\times10^{-5}$&$8.94\times10^{-6}$&$8.96\times10^{-6}$&$9.24\times10^{-6}$&$1.26\times10^{-5}$\\
\hline
\end{tabular}
\end{center}
\caption{Training and testing performance for aneurysm case with varying viscosity/geometry}
\label{tab:apptab2}
}
\end{table}

\clearpage
\section*{Appendix B}
\label{sec:ap2}
\begin{figure}[htp]
	\centering 
	\subfloat[Swish-$\beta$, hard]{\includegraphics[width=0.2\textwidth]{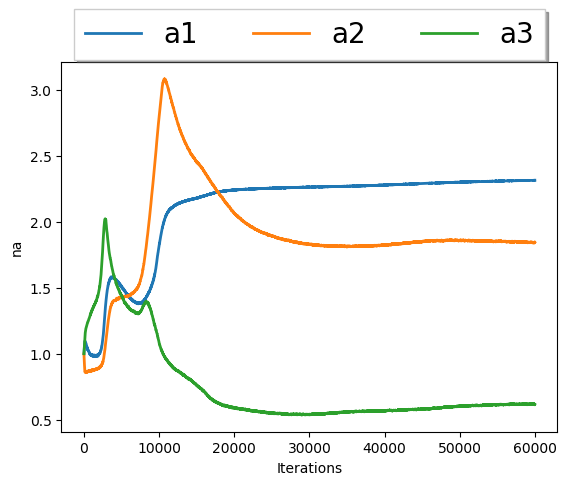}}
	\subfloat[Tanh-$n1$, hard]{\includegraphics[width=0.2\textwidth]{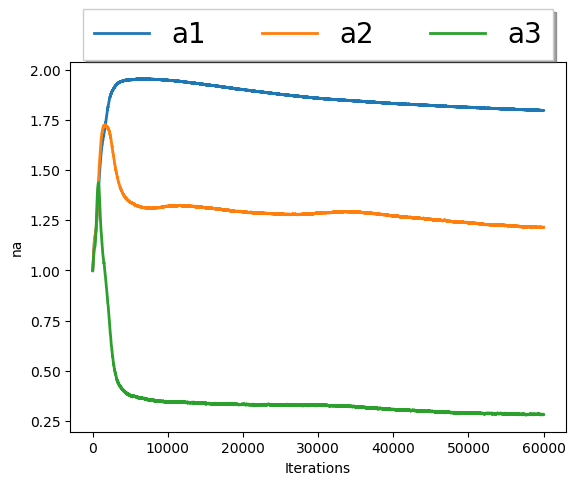}}
	\subfloat[Tanh-$n5$, hard]{\includegraphics[width=0.2\textwidth]{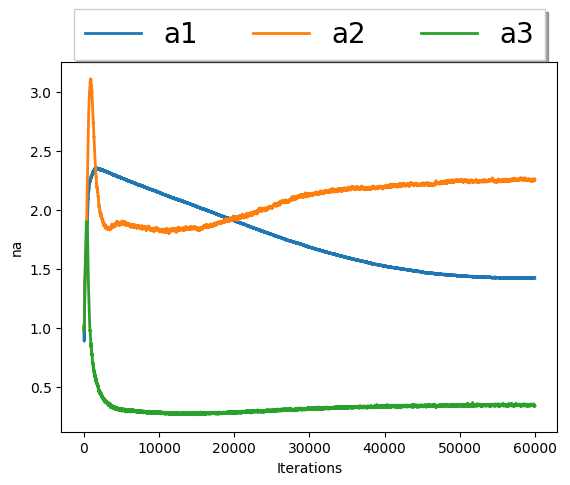}}
	\subfloat[Tanh-$n10$, hard,]{\includegraphics[width=0.2\textwidth]{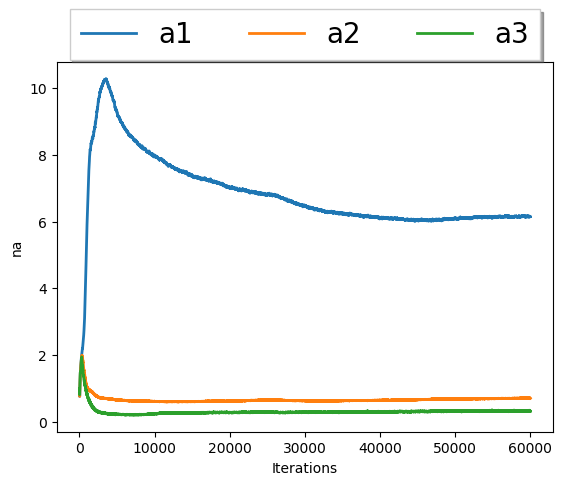}}
	\subfloat[Tanh-$n20$, hard]{\includegraphics[width=0.2\textwidth]{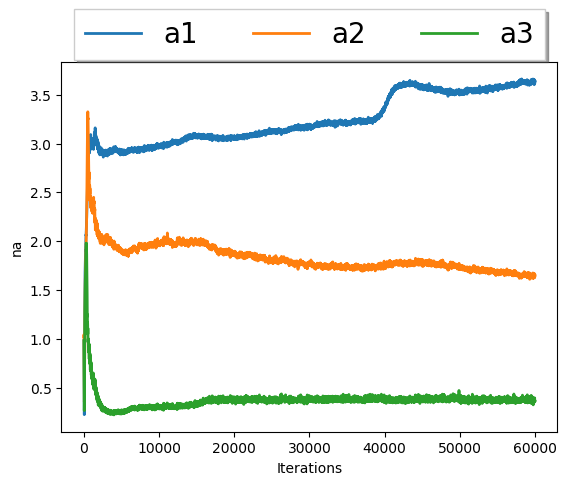}}
	\vfill
	\subfloat[Swish-$\beta$, soft]{\includegraphics[width=0.2\textwidth]{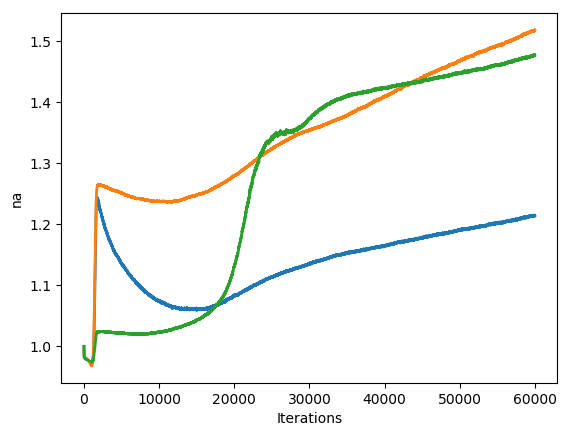}}
	\subfloat[Tanh-$n1$, soft]{\includegraphics[width=0.2\textwidth]{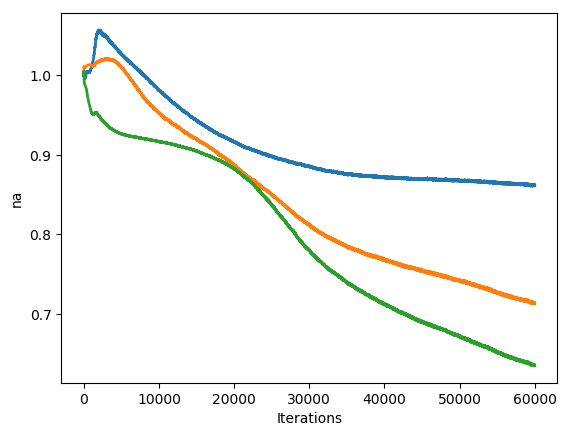}}
	\subfloat[Tanh-$n5$, soft]{\includegraphics[width=0.2\textwidth]{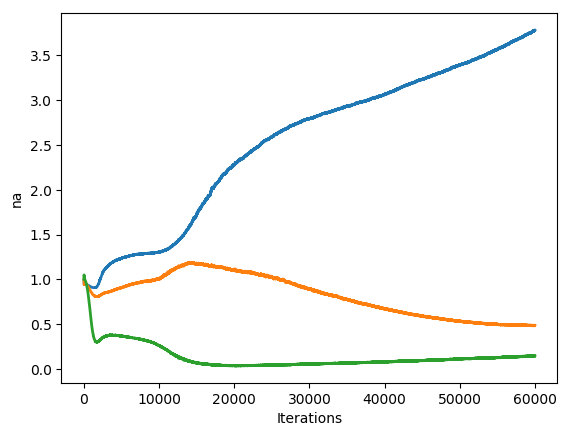}}
	\subfloat[Tanh-$n10$, soft]{\includegraphics[width=0.2\textwidth]{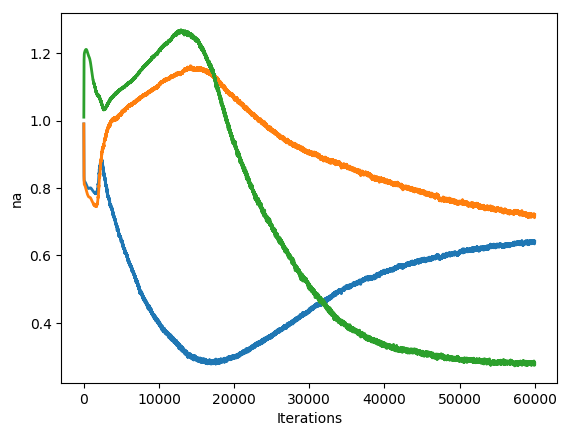}}
	\subfloat[Tanh-$n20$, soft]{\includegraphics[width=0.2\textwidth]{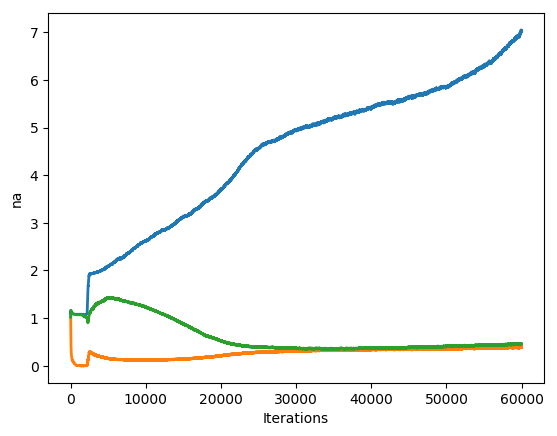}}
	\caption{Convergence histories of trainable parameters ($\beta$ for Swish and $na$ for Tahn) of adaptive activation functions, where panels (a)--(e) are for cases with hard BC enforcement and panels (f)--(j) are for cases with soft BC enforcement. The legend $a1, a2, a3$ refer to three different layers.}
    \label{fig:adpAc}
\end{figure}

Figure~\ref{fig:adpAc} shows the convergence histories of trainable parameters of different adaptive activation functions. The comparison is also made between cases where the BCs are imposed in hard and soft manners. It can be found that the parameters of activation function are more likely to converge with hard BC enforcement (panels (a)--(e)) compared to the cases with soft BC enforcement (panels (f)--(j)). 


\begin{thebibliography}{10}
	\expandafter\ifx\csname url\endcsname\relax
	\def\url#1{\texttt{#1}}\fi
	\expandafter\ifx\csname urlprefix\endcsname\relax\def\urlprefix{URL }\fi
	\expandafter\ifx\csname href\endcsname\relax
	\def\href#1#2{#2} \def\path#1{#1}\fi
	
	\bibitem{anderson1995computational}
	J.~D. Anderson, J.~Wendt, Computational fluid dynamics, Vol. 206, Springer,
	1995.
	
	\bibitem{benner2015survey}
	P.~Benner, S.~Gugercin, K.~Willcox, A survey of projection-based model
	reduction methods for parametric dynamical systems, SIAM Review 57~(4) (2015)
	483--531.
	
	\bibitem{benner2017model}
	P.~Benner, A.~Cohen, M.~Ohlberger, K.~Willcox, Model reduction and
	approximation: theory and algorithms, Vol.~15, SIAM, 2017.
	
	\bibitem{lassila2014model}
	T.~Lassila, A.~Manzoni, A.~Quarteroni, G.~Rozza, Model order reduction in fluid
	dynamics: challenges and perspectives, in: Reduced Order Methods for modeling
	and computational reduction, Springer, 2014, pp. 235--273.
	
	\bibitem{huang2018challenges}
	C.~Huang, K.~Duraisamy, C.~Merkle, Challenges in reduced order modeling of
	reacting flows, in: 2018 Joint Propulsion Conference, 2018, p. 4675.
	
	\bibitem{chaturantabut2010nonlinear}
	S.~Chaturantabut, D.~C. Sorensen, Nonlinear model reduction via discrete
	empirical interpolation, SIAM Journal on Scientific Computing 32~(5) (2010)
	2737--2764.
	
	\bibitem{peherstorfer2018model}
	B.~Peherstorfer, Model reduction for transport-dominated problems via online
	adaptive bases and adaptive sampling, arXiv preprint arXiv:1812.02094.
	
	\bibitem{peherstorfer2018stabilizing}
	B.~Peherstorfer, Z.~Drma{\v{c}}, S.~Gugercin, Stabilizing discrete empirical
	interpolation via randomized and deterministic oversampling, arXiv preprint
	arXiv:1808.10473.
	
	\bibitem{ren2010finite}
	W.-X. Ren, H.-B. Chen, Finite element model updating in structural dynamics by
	using the response surface method, Engineering Structures 32~(8) (2010)
	2455--2465.
	
	\bibitem{regis2007stochastic}
	R.~G. Regis, C.~A. Shoemaker, A stochastic radial basis function method for the
	global optimization of expensive functions, INFORMS Journal on Computing
	19~(4) (2007) 497--509.
	
	\bibitem{kennedy2000predicting}
	M.~C. Kennedy, A.~O'Hagan, Predicting the output from a complex computer code
	when fast approximations are available, Biometrika 87~(1) (2000) 1--13.
	
	\bibitem{atkinson2019structured}
	S.~Atkinson, N.~Zabaras, Structured bayesian gaussian process latent variable
	model: Applications to data-driven dimensionality reduction and
	high-dimensional inversion, Journal of Computational Physics 383 (2019)
	166--195.
	
	\bibitem{xiu2002wiener}
	D.~Xiu, G.~E. Karniadakis, The wiener--askey polynomial chaos for stochastic
	differential equations, SIAM Journal on Scientific Computing 24~(2) (2002)
	619--644.
	
	\bibitem{najm2009uncertainty}
	H.~N. Najm, Uncertainty quantification and polynomial chaos techniques in
	computational fluid dynamics, Annual Review of Fluid Mechanics 41 (2009)
	35--52.
	
	\bibitem{yang2017general}
	X.~Yang, X.~Wan, L.~Lin, L.~Huan, A general framework for enhancing sparsity of
	generalized polynomial chaos expansions, International Journal for
	Uncertainty Quantification 9~(3) (2017) 221--243.
	
	\bibitem{le2010spectral}
	O.~Le~Ma{\^\i}tre, O.~M. Knio, Spectral methods for uncertainty quantification:
	with applications to computational fluid dynamics, Springer Science \&
	Business Media, 2010.
	
	\bibitem{wang2018propagation}
	J.-X. Wang, C.~J. Roy, H.~Xiao, Propagation of input uncertainty in presence of
	model-form uncertainty: a multifidelity approach for computational fluid
	dynamics applications, ASCE-ASME Journal of Risk and Uncertainty in
	Engineering Systems, Part B: Mechanical Engineering 4~(1) (2018) 011002.
	
	\bibitem{zhu2018bayesian}
	Y.~Zhu, N.~Zabaras, Bayesian deep convolutional encoder--decoder networks for
	surrogate modeling and uncertainty quantification, Journal of Computational
	Physics 366 (2018) 415--447.
	
	\bibitem{tripathy2018deep}
	R.~K. Tripathy, I.~Bilionis, Deep uq: Learning deep neural network surrogate
	models for high dimensional uncertainty quantification, Journal of
	Computational Physics 375 (2018) 565--588.
	
	\bibitem{mo2018deep}
	S.~Mo, N.~Zabaras, X.~Shi, J.~Wu, Deep autoregressive neural networks for
	high-dimensional inverse problems in groundwater contaminant source
	identification, Water Resources Research.
	
	\bibitem{scarselli1998universal}
	F.~Scarselli, A.~C. Tsoi, Universal approximation using feedforward neural
	networks: A survey of some existing methods, and some new results, Neural
	networks 11~(1) (1998) 15--37.
	
	\bibitem{hutzenthaler2018overcoming}
	M.~Hutzenthaler, A.~Jentzen, T.~Kruse, T.~A. Nguyen, P.~von Wurstemberger,
	Overcoming the curse of dimensionality in the numerical approximation of
	semilinear parabolic partial differential equations, arXiv preprint
	arXiv:1807.01212.
	
	\bibitem{grohs2018proof}
	P.~Grohs, F.~Hornung, A.~Jentzen, P.~Von~Wurstemberger, A proof that artificial
	neural networks overcome the curse of dimensionality in the numerical
	approximation of black-scholes partial differential equations, arXiv preprint
	arXiv:1809.02362.
	
	\bibitem{hutzenthaler2019proof}
	M.~Hutzenthaler, A.~Jentzen, T.~Kruse, T.~A. Nguyen, A proof that rectified
	deep neural networks overcome the curse of dimensionality in the numerical
	approximation of semilinear heat equations, arXiv preprint arXiv:1901.10854.
	
	\bibitem{lee2018basic}
	S.~Lee, N.~Baker, Basic research needs for scientific machine learning: Core
	technologies for artificial intelligence, Tech. rep., USDOE Office of Science
	(SC)(United States) (2018).
	
	\bibitem{carleo2017solving}
	G.~Carleo, M.~Troyer, Solving the quantum many-body problem with artificial
	neural networks, Science 355~(6325) (2017) 602--606.
	
	\bibitem{wang2017physics}
	J.-X. Wang, J.-L. Wu, H.~Xiao, Physics-informed machine learning approach for
	reconstructing reynolds stress modeling discrepancies based on dns data,
	Physical Review Fluids 2~(3) (2017) 034603.
	
	\bibitem{ling2016machine}
	J.~Ling, R.~Jones, J.~Templeton, Machine learning strategies for systems with
	invariance properties, Journal of Computational Physics 318 (2016) 22--35.
	
	\bibitem{shanahan2018machine}
	P.~E. Shanahan, D.~Trewartha, W.~Detmold, Machine learning action parameters in
	lattice quantum chromodynamics, Physical Review D 97~(9) (2018) 094506.
	
	\bibitem{king2018deep}
	R.~King, O.~Hennigh, A.~Mohan, M.~Chertkov, From deep to physics-informed
	learning of turbulence: Diagnostics, arXiv preprint arXiv:1810.07785.
	
	\bibitem{brunton2019data}
	S.~L. Brunton, J.~N. Kutz, Data-driven Science and Engineering: Machine
	Learning, Dynamical Systems, and Control, Cambridge University Press, 2019.
	
	\bibitem{lecun2015deep}
	Y.~LeCun, Y.~Bengio, G.~Hinton, Deep learning, Nature 521~(7553) (2015) 436.
	
	\bibitem{raissi2019physics}
	M.~Raissi, P.~Perdikaris, G.~Karniadakis, Physics-informed neural networks: A
	deep learning framework for solving forward and inverse problems involving
	nonlinear partial differential equations, Journal of Computational Physics
	378 (2019) 686--707.
	
	\bibitem{hannun2014deep}
	A.~Hannun, C.~Case, J.~Casper, B.~Catanzaro, G.~Diamos, E.~Elsen, R.~Prenger,
	S.~Satheesh, S.~Sengupta, A.~Coates, et~al., Deep speech: Scaling up
	end-to-end speech recognition, arXiv preprint arXiv:1412.5567.
	
	\bibitem{lee1990neural}
	H.~Lee, I.~S. Kang, Neural algorithm for solving differential equations,
	Journal of Computational Physics 91~(1) (1990) 110--131.
	
	\bibitem{lagaris1998artificial}
	I.~E. Lagaris, A.~Likas, D.~I. Fotiadis, Artificial neural networks for solving
	ordinary and partial differential equations, IEEE Transactions on Nneural
	Networks 9~(5) (1998) 987--1000.
	
	\bibitem{lagaris2000neural}
	I.~E. Lagaris, A.~C. Likas, D.~G. Papageorgiou, Neural-network methods for
	boundary value problems with irregular boundaries, IEEE Transactions on
	Neural Networks 11~(5) (2000) 1041--1049.
	
	\bibitem{raissi2018hidden}
	M.~Raissi, A.~Yazdani, G.~E. Karniadakis, Hidden fluid mechanics: A
	navier-stokes informed deep learning framework for assimilating flow
	visualization data, arXiv preprint arXiv:1808.04327.
	
	\bibitem{raissi2019deep}
	M.~Raissi, Z.~Wang, M.~S. Triantafyllou, G.~E. Karniadakis, Deep learning of
	vortex-induced vibrations, Journal of Fluid Mechanics 861 (2019) 119--137.
	
	\bibitem{tartakovsky2018learning}
	A.~M. Tartakovsky, C.~O. Marrero, D.~Tartakovsky, D.~Barajas-Solano, Learning
	parameters and constitutive relationships with physics informed deep neural
	networks, arXiv preprint arXiv:1808.03398.
	
	\bibitem{meng2019composite}
	X.~Meng, G.~E. Karniadakis, A composite neural network that learns from
	multi-fidelity data: Application to function approximation and inverse pde
	problems, arXiv preprint arXiv:1903.00104.
	
	\bibitem{zhang2018quantifying}
	D.~Zhang, L.~Lu, L.~Guo, G.~E. Karniadakis, Quantifying total uncertainty in
	physics-informed neural networks for solving forward and inverse stochastic
	problems, arXiv preprint arXiv:1809.08327.
	
	\bibitem{yang2018adversarial}
	Y.~Yang, P.~Perdikaris, Adversarial uncertainty quantification in
	physics-informed neural networks, arXiv preprint arXiv:1811.04026.
	
	\bibitem{sharma2018weakly}
	R.~Sharma, A.~B. Farimani, J.~Gomes, P.~Eastman, V.~Pande, Weakly-supervised
	deep learning of heat transport via physics informed loss, arXiv preprint
	arXiv:1807.11374.
	
	\bibitem{nabian2018physics}
	M.~A. Nabian, H.~Meidani, Physics-informed regularization of deep neural
	networks, arXiv preprint arXiv:1810.05547.
	
	\bibitem{xu2019neural}
	K.~Xu, E.~Darve, The neural network approach to inverse problems in
	differential equations, arXiv preprint arXiv:1901.07758.
	
	\bibitem{holland2019towards}
	J.~R. Holland, J.~D. Baeder, K.~Duraisamy, Towards integrated field inversion
	and machine learning with embedded neural networks for rans modeling, in:
	AIAA Scitech 2019 Forum, 2019, p. 1884.
	
	\bibitem{stewart2017label}
	R.~Stewart, S.~Ermon, Label-free supervision of neural networks with physics
	and domain knowledge, in: Thirty-First AAAI Conference on Artificial
	Intelligence, 2017, pp. 2576--2582.
	
	\bibitem{sirignano2018dgm}
	J.~Sirignano, K.~Spiliopoulos, Dgm: A deep learning algorithm for solving
	partial differential equations, Journal of Computational Physics 375 (2018)
	1339--1364.
	
	\bibitem{berg2018unified}
	J.~Berg, K.~Nystr{\"o}m, A unified deep artificial neural network approach to
	partial differential equations in complex geometries, Neurocomputing 317
	(2018) 28--41.
	
	\bibitem{weinan2017deep}
	E.~Weinan, J.~Han, A.~Jentzen, Deep learning-based numerical methods for
	high-dimensional parabolic partial differential equations and backward
	stochastic differential equations, Communications in Mathematics and
	Statistics 5~(4) (2017) 349--380.
	
	\bibitem{beck2017machine}
	C.~Beck, E.~Weinan, A.~Jentzen, Machine learning approximation algorithms for
	high-dimensional fully nonlinear partial differential equations and
	second-order backward stochastic differential equations, Journal of Nonlinear
	Science (2017) 1--57.
	
	\bibitem{beck2018solving}
	C.~Beck, S.~Becker, P.~Grohs, N.~Jaafari, A.~Jentzen, Solving stochastic
	differential equations and {Kolmogorov} equations by means of deep learning,
	arXiv preprint arXiv:1806.00421.
	
	\bibitem{weinan2018deep}
	E.~Weinan, B.~Yu, The deep ritz method: A deep learning-based numerical
	algorithm for solving variational problems, Communications in Mathematics and
	Statistics 6~(1) (2018) 1--12.
	
	\bibitem{han2018solving}
	J.~Han, A.~Jentzen, E.~Weinan, Solving high-dimensional partial differential
	equations using deep learning, Proceedings of the National Academy of
	Sciences 115~(34) (2018) 8505--8510.
	
	\bibitem{nabian2018deep}
	M.~A. Nabian, H.~Meidani, A deep neural network surrogate for high-dimensional
	random partial differential equations, arXiv preprint arXiv:1806.02957.
	
	\bibitem{karumuri2019simulator}
	S.~Karumuri, R.~Tripathy, I.~Bilionis, J.~Panchal, Simulator-free solution of
	high-dimensional stochastic elliptic partial differential equations using
	deep neural networks, arXiv preprint arXiv:1902.05200.
	
	\bibitem{zhu2019physics}
	Y.~Zhu, N.~Zabaras, P.-S. Koutsourelakis, P.~Perdikaris, Physics-constrained
	deep learning for high-dimensional surrogate modeling and uncertainty
	quantification without labeled data, arXiv preprint arXiv:1901.06314.
	
	\bibitem{baydin2018automatic}
	A.~G. Baydin, B.~A. Pearlmutter, A.~A. Radul, J.~M. Siskind, Automatic
	differentiation in machine learning: a survey, Journal of Machine Learning
	Research 18 (2018) 1--43.
	
	\bibitem{paszke2017automatic}
	A.~Paszke, S.~Gross, S.~Chintala, G.~Chanan, E.~Yang, Z.~DeVito, Z.~Lin,
	A.~Desmaison, L.~Antiga, A.~Lerer, Automatic differentiation in {PyTorch},
	in: NIPS Autodiff Workshop, 2017.
	
	\bibitem{abadi2016tensorflow}
	M.~Abadi, P.~Barham, J.~Chen, Z.~Chen, A.~Davis, J.~Dean, M.~Devin,
	S.~Ghemawat, G.~Irving, M.~Isard, et~al., Tensorflow: A system for
	large-scale machine learning, in: 12th USENIX Symposium on Operating Systems
	Design and Implementation (OSDI 16), 2016, pp. 265--283.
	
	\bibitem{bastien2012theano}
	F.~Bastien, P.~Lamblin, R.~Pascanu, J.~Bergstra, I.~J. Goodfellow, A.~Bergeron,
	N.~Bouchard, Y.~Bengio, Theano: new features and speed improvements. deep
	learning and unsupervised feature learning, in: Neural Information Processing
	Systems Workshop (NIPS), 2012, pp. 1--10.
	
	\bibitem{kleinberg2018alternative}
	R.~Kleinberg, Y.~Li, Y.~Yuan, An alternative view: When does {SGD} escape local
	minima?, arXiv preprint arXiv:1802.06175.
	
	\bibitem{marquez2017imposing}
	P.~M{\'a}rquez-Neila, M.~Salzmann, P.~Fua, Imposing hard constraints on deep
	networks: Promises and limitations, arXiv preprint arXiv:1706.02025.
	
	\bibitem{ramachandran2017swish}
	P.~Ramachandran, B.~Zoph, Q.~V. Le, Swish: a self-gated activation function,
	arXiv preprint arXiv:1710.05941 7.
	
	\bibitem{kingma2014adam}
	D.~P. Kingma, J.~Ba, Adam: A method for stochastic optimization, arXiv preprint
	arXiv:1412.6980.
	
	\bibitem{Glorot10understandingthe}
	X.~Glorot, Y.~Bengio, Understanding the difficulty of training deep feedforward
	neural networks, in: In Proceedings of the International Conference on
	Artificial Intelligence and Statistics (AISTATS’10). Society for Artificial
	Intelligence and Statistics, 2010.
	
	\bibitem{he2015delving}
	K.~He, X.~Zhang, S.~Ren, J.~Sun, Delving deep into rectifiers: Surpassing
	human-level performance on imagenet classification, in: Proceedings of the
	IEEE International Conference on Computer Vision, 2015, pp. 1026--1034.
	
	\bibitem{Jasak07openfoam:a}
	H.~Jasak, A.~Jemcov, U.~Kingdom, {OpenFOAM}: A {C++} library for complex
	physics simulations, in: International Workshop on Coupled Methods in
	Numerical Dynamics, IUC, 2007, pp. 1--20.
	
	\bibitem{berger2000flows}
	S.~Berger, L.-D. Jou, Flows in stenotic vessels, Annual review of fluid
	mechanics 32~(1) (2000) 347--382.
	
	\bibitem{brisman2006cerebral}
	J.~L. Brisman, J.~K. Song, D.~W. Newell, Cerebral aneurysms, New England
	journal of medicine 355~(9) (2006) 928--939.
	
	\bibitem{cebral2011association}
	J.~R. Cebral, F.~Mut, J.~Weir, C.~M. Putman, Association of hemodynamic
	characteristics and cerebral aneurysm rupture, American Journal of
	Neuroradiology 32~(2) (2011) 264--270.
	
	\bibitem{chalouhi2013review}
	N.~Chalouhi, B.~L. Hoh, D.~Hasan, Review of cerebral aneurysm formation,
	growth, and rupture, Stroke 44~(12) (2013) 3613--3622.
	
	\bibitem{maas2013rectifier}
	A.~L. Maas, A.~Y. Hannun, A.~Y. Ng, Rectifier nonlinearities improve neural
	network acoustic models, in: Proceedings of the 30th Annual International
	Conference on Machine Learning, Vol.~30, 2013, p.~3.
	
	\bibitem{ramachandran2017searching}
	P.~Ramachandran, B.~Zoph, Q.~V. Le, Searching for activation functions, arXiv
	preprint arXiv:1710.05941.
	
	\bibitem{jagtap2019adaptive}
	A.~D. Jagtap, G.~E. Karniadakis, Adaptive activation functions accelerate
	convergence in deep and physics-informed neural networks, arXiv preprint
	arXiv:1906.01170.
	
	\bibitem{hamerly2019large}
	R.~Hamerly, L.~Bernstein, A.~Sludds, M.~Solja{\v{c}}i{\'c}, D.~Englund,
	Large-scale optical neural networks based on photoelectric multiplication,
	Physical Review X 9~(2) (2019) 021032.
	
\end{thebibliography}

\end{document}